\documentclass[prc,superscriptaddress,nofootinbib,noeprint,dvipsnames,aps,10pt]{revtex4-2}

\usepackage[english]{babel}
\usepackage[hidelinks]{hyperref}
\makeatletter\AtBeginDocument{\let\@elt\relax}\makeatother

\usepackage{booktabs}
\usepackage{longtable}
\usepackage{tabularx}
\usepackage{graphicx}
\usepackage{amsmath}
\usepackage{float}
\usepackage[english,capitalise]{cleveref}
\usepackage{mathtools}
\usepackage{bbm}

\usepackage[utf8]{inputenc}

\usepackage[dvipsnames]{xcolor}

\usepackage{physics}
\usepackage{csquotes}
\usepackage{multirow}
\usepackage{ifthen}
\usepackage[normalem]{ulem}
\usepackage{orcidlink}

\usepackage{relsize}

\makeindex

\usepackage{our_newcommands0722}

\newcommand{\tof}[4]{\,{}^{#1}\mathlarger{\mathcal{T}}^{\,#2|#3}_{#4}}

\newcommand{\moop}{\Omega}

\renewcommand{\vec}[1]{\v{#1}}
\newcommand{\lmi}{l_{\mathrm{max}}^{(i)}}
\newcommand{\lmf}{l_{\mathrm{max}}^{(f)}}
\newcommand{\Lt}{\tilde{L}}

\newcommand{\kjipv}[1]{\v{\kappa}_{jip}{\K{#1}}}
\newcommand{\kjiqv}[1]{\v{\kappa}_{jiq}{\K{#1}}}
\newcommand{\kjip}[1]{\kappa_{jip}{\K{#1}}}
\newcommand{\kjiq}[1]{\kappa_{jiq}{\K{#1}}}

\begin{document}

\title{\texorpdfstring{Finite-range EFT for the \(E1\) strength distribution of \(^6\)He}{Finite-range EFT for the E1 strength distribution of 6He}}

\author{Matthias G\"{o}bel \orcidlink{0000-0002-7232-0033}}
\email[E-mail: ]{goebel.matthias@ujf.cas.cz}
\affiliation{Nuclear Physics Institute, Czech Academy of Sciences, 25068 \scv{R}e\scv{z}, Czech Republic}
\affiliation{Istituto Nazionale di Fisica Nucleare, Sezione di Pisa, 56127 Pisa, Italy}

\author{Hans-Werner Hammer \orcidlink{0000-0002-2318-0644}}
\affiliation{Technische Universit\"{a}t Darmstadt, Department of Physics, Institut f\"{u}r Kernphysik, 64289 Darmstadt, Germany}
\affiliation{ExtreMe Matter Institute EMMI and Helmholtz Forschungsakademie Hessen für FAIR (HFHF), GSI Helmholtzzentrum f\"{u}r Schwerionenforschung GmbH, 64291 Darmstadt, Germany}

\author{Daniel R. Phillips \orcidlink{0000-0003-1596-9087}}
\affiliation{Institute of Nuclear and Particle Physics and Department of Physics and Astronomy, Ohio University, Athens, OH 45701, USA}
\affiliation{Department of Physics, Chalmers University of Technology, SE-41296 G\"oteborg, Sweden}


\date{\today}

\begin{abstract}
Halo effective field theory (Halo EFT) is a powerful tool to describe halo nuclei and predict low-energy observables with quantified uncertainties.
However, in the case that there is a leading-order interaction determined by two or more effective-range parameters, such as the \(^2P_{3/2}\) \(n\alpha\) interaction in \(^6\)He, 
the standard implementation in the dimer formalism leads to 
an energy-dependent interaction. This complicates the 
construction of a Hilbert space of states, especially beyond the two-body problem. 
As an alternative, we propose the use of a finite-range formulation of Halo EFT, which avoids these complications. 
For definiteness, we use separable interactions with Yamaguchi-like
form factors, but other choices are possible.
We solve for the ${}^6$He bound state in this finite-range EFT up to next-to-leading order (NLO) in the Halo EFT power counting and calculate the ground-state \(E1\) strength distribution of \(^6\)He at this order. The shape of the resulting distribution agrees with that obtained in the dimer formalism of the EFT, but finite-range EFT does not require the use of a non-standard wave function normalization condition. 
We also calculate the root-mean-square charge radius of \(^6\)He and find $2.06 \pm 0.35$~fm at LO and $2.00 \pm 0.09$~fm at NLO, in agreement with experimental data.
To calculate the full \(E1\) strength distribution final-state interactions must be incorporated.
We approximate the full-three-body scattering operator first by single M{\o}ller operators and then by products of up to three M{\o}ller operators. 
The resulting NLO \(E1\) strength distribution agrees with the experimental data within theory uncertainties.
\end{abstract}
\maketitle


\section{Introduction}
\label{sec:intro}

Nuclei near the edge of stability are typically unstable to weak decay. But the advent of 
rare-isotope-beam facilities makes them amenable to experimental investigation. 
Neutron-rich systems near the edge of stability frequently form ``neutron halos'': they consist
of a compact, tightly bound core and
a few loosely bound halo neutrons \cite{Zhukov:1993aw,Hansen:1995pu,Jonson:2004,Riisager:2012it}.
The halo neutrons spend the majority of their time outside the range
of the interaction, i.e., in the classically forbidden region. This extreme quantum character of halo nuclei makes them a fertile ground for investigations based on few-body methods, where these quantal aspects can be handled, in principle, exactly.

A powerful framework for describing halo nuclei is halo
effective field theory (Halo EFT) \cite{Bertulani:2002sz,Bedaque:2003wa},
which is obtained by complementing nuclear cluster models with a
systematic power counting scheme (see Refs.~\cite{Hammer:2017tjm,Hammer:2019poc,Hammer:2022lhx} for reviews).
Halo EFT exploits the separation of scales between
the tightly bound core and the loosely bound valence
neutrons to  predict observables in an expansion in the ratio of these scales.
This scheme provides a systematically improvable description of halo nuclei.
Uncertainties due to higher-order effects can be estimated from the putative expansion parameter and the convergence pattern 
at the orders that have been calculated.

However, in higher partial waves the standard implementation of Halo EFT
with dimer fields comes with practical difficulties.
To explain those, we start with the \(s\)-wave case where these
difficulties are not present.
Consider an \(s\)-wave interaction with an unnaturally large scattering
length \(a_0\). This corresponds to a shallow bound or virtual state
in this interaction channel.
According to the standard power counting, the leading-order (LO) on-shell
t-matrix 
is determined by just one parameter, the scattering length  \(a_0\).
The coupling of a contact interaction in the Lagrangian can be tuned such
that the t-matrix in the scattering length approximation
can be reproduced exactly.
Some higher-order terms in the effective range expansion of the t-matrix
are also induced, but they can be made arbitraily small by increasing the
cutoff.

If the power counting stipulates that the leading-order t-matrix has
two or more parameters,  as is the case for \(p\)-waves, the situation is more
complicated.
A typical example is the \(n\alpha\) interaction in
the \(^2P_{3/2}\) partial wave, giving rise to a shallow \(^5\)He resonance.
For this system, two power counting scenarios have been proposed
\cite{Bertulani:2002sz,Bedaque:2003wa}, but the difficulty exists for
both scenarios.
In either the LO t-matrix is determined by two effective range parameters,
the scattering volume $a_1$ and the effective range $r_1$.
It is possible to represent this scenario with contact interactions in the
dimer formalism.
However, this requires a kinetic term for the dimer, such that the dimer
becomes dynamic. 
In the language of quantum mechanics, this interaction corresponds to an
energy-dependent potential requiring modifications to the standard
quantum mechanics framework as explained in
Refs.~\cite{McKellar:1983muf,formanek04}.
In particular, it requires a non-standard normalization of wave functions.
These modifications introduce correction terms in the evaluation of
expectation values, which correspond to counter terms in the EFT
picture.

A Halo EFT treatment for \(^6\)He was developed by Ji et al.~\cite{Ji:2014wta}.
This treatment was extended to wave functions and probability densities in Ref.~\cite{Gobel:2019jba}. In this paper, the correction terms arising from the non-standard normalization of the wave function
were explored for the probability density.
An approximate implementation has been shown
to work well in this case.
While the influence of the correction terms was small in the low-momentum components of the wave function,
it becomes considerable at higher momenta.
Because of the normalization condition the correction terms influence even low-energy observables; but the shape of low-energy observables is not affected by this issue.

In this paper, we propose an alternative realization of Halo EFT which
avoids the difficulty of handling energy-dependent interactions.
Instead, we employ separable finite-range interactions. Here we implement the EFT using Yamaguchi potentials and variations thereof. We recapitulate previous results which show that these interactions can be used to reproduce the $nc$ $t$-matrix
at leading order (LO) and next-to-leading order (NLO). The difference between the LO and NLO results is of the size expected based on the
EFT power counting. 

Yamaguchi potentials are separable potentials of rank one with
so-called Yamaguchi form factors (see Sec.~\ref{sec:found} for explicit
expressions).
They are typically specified in momentum space and were originally
introduced in Ref.~\cite{Yamaguchi:1954mp}. 
The standard Yamaguchi interactions have two parameters, a coupling
constant and a range parameter in the form factor.
Thus they can straightforwardly be applied to power counting scenarios
which require two effective range parameters at LO without introducing
any energy dependent interactions.
If only one ERE parameter is given, one can either resort to a contact
interaction or use a Yamaguchi interaction and vary the undetermined parameter
of the potential to get a span of possible results.
For interactions determined by more than two ERE parameters, e.g., those
needed at higher orders, a modified version of the Yamaguchi
potential with more complicated forms and more parameters can be used. 

While we use separable Yamaguchi interactions for convenience, other finite-range
interactions characterized by a coupling constant and a range would
work equally well. 
We note that there has been particular recent interest in separable interactions with a square-root form factor. 
This form factor can be chosen so that the  $s$-wave effective range and scattering length are both reproduced at LO in the EFT, while all other effective range parameters vanish exactly.
Beane and Farrell argue that the resulting S-matrix then possess a UV/IR symmetry~\cite{Beane:2021dab}. When employed to regulate the 
short-range piece of the nucleon-nucleon interaction in chiral EFT the same form factor
improves the convergence of that EFT in the ${}^1$S$_0$ channel~\cite{Peng:2021pvo}.

We note that the use of finite range interactions to describe
low-energy nuclear properties has a long history (see, e.g., Ref.~\cite{Reid:1968sq} for a review). In meson-exchange models of the nucleon-nucleon interaction \cite{Yukawa:1935xg} and in chiral effective field theory \cite{Weinberg:1990rz,Weinberg:1991um}, finite-range forces arise naturally from the exchange of pions. The tail from one-pion exchange is clearly seen in partial wave analyses of the nucleon-nucleon interaction \cite{Stoks:1993tb}.
Since our focus here is not on finite-range interactions in chiral EFT, we refrain from discussing the vast literature on this topic (see, e.g., Refs.~\cite{Epelbaum:2008ga,Phillips:2016mov,Hammer:2019poc,Machleidt:2024bwl} for reviews) and discuss only recent related work in short-range EFTs.
In this context, finite-range interactions have also been considered for very low-energy effective interactions, where the finite range of nuclear forces is not resolved.
Contessi {\it et al.}~\cite{Contessi:2023yoz} argued that
finite-range interactions could be used to partially resum
the effective range in the leading order $s$-wave amplitude, thereby
improving the rate of convergence of the EFT without altering its radius
of convergence~\cite{Contessi:2024vae}. They have also been used to examine the consequences of large-scattering-length universality for few-body systems in Ref.~\cite{Kievsky:2021ghz}.

An implementation of finite-range interactions in a non-local version of Halo EFT was  explored in the context 
of $^6$He in Ref.~\cite{Li:2023hwe}. This work showed that
the features of the ${}^6$He bound-state problem
encountered in the 
standard Halo EFT treatment of Ref.~\cite{Ji:2014wta} could be 
recovered in such an implementation. 
More recently, Pinilla {\it et al.}~\cite{Pinilla:2024xwz}, calculated the ground state energy, root mean square radius, and $E1$ strength distribution in a three-body model of $^6$He derived from cluster effective field theory. They used a non-local configuration space representation of the cluster interactions in hyperspherical coordinates and found good agreement for the
mean square radius. 
There is also a recent Halo effective theory calculation of the \(E1\) strength of \({}^6\)He by Bertulani which uses asymptotic wave functions in the hyperspherical harmonics formalism \cite{Bertulani:2023hia}. The \(^1S_0\) nn interaction and the \(^2P_{3/2}\) nc interaction are taken into account.

The insights these EFT calculations provide on \(^6\)He build on earlier work in three-body ($\alpha nn$) models; 
such studies have a history of more than fifty years.
Some used a momentum-space (Faddeev) formulation in combination with separable interactions (see, e.g., Refs.~\cite{Hebach:1967bpg,Ghovanlou:1974zza}), others used a coordinate-space formulation based on the hyperspherical harmonics formalism with local interactions (see, e.g, Refs.~\cite{Chulkov:1990ac,Zhukov:1993aw}). In more recent years such hyperpherical-harmonics calculations have been extended to the \(^7\)He and \(^8\)He systems. For example, Ref.~\cite{Higgins:2024lpf} constructed simple $\alpha n$ and $\alpha nn$ potentials and then computed \(^6\)He bound-state and resonance properties, scattering states of the \(^7\)He system, and properties of the \(^8\)He nucleus in its ground state. 

The \(E1\) strength distribution
determines the Coulomb dissociation cross section of \(^6\)He
and has been measured experimentally \cite{Aumann:1999mb,Wang:2002sm,Sun:2021efo}.
The \(E1\) breakup spectrum of \(^6\)He has been calculated in a number of cluster models, e.g., Refs.~\cite{Cobis:1998gx,Forssen:2001rj,Grigorenko:2020jts,Kikuchi:2010zzb}.
Detailed investigations of the influence of initial-state correlations and final-state interactions on the \(E1\) breakup can be found, e.g., in Refs.~\cite{Grigorenko:2020jts,Kikuchi:2010zzb}.
Final-state interactions
were found to play a crucial role in the \(^6\)He \(E1\) spectrum, 
with
\(nn\) and \(\alpha n\) final-state interactions having approximately equal importance~\cite{Kikuchi:2010zzb}. The \(E1\) breakup of \(^6\)He thus provides an experimental manifestation of dineutron clustering: the impact of this effect on the \(E1\) breakup of two-neutron halo nuclei is emphasized in the the recent review of Nakamura {\it et al.}~\cite{Nakamura:2026jnq}. 

In the present investigation we calculate the low-energy
$E1$ response of $^6$He in finite-range Halo EFT up to next-to-leading order in the EFT expansion.
At leading order the two-neutron halo nucleus \(^6\)He features a
two-parameter \(^2P_{3/2}\) \(n\alpha\) interaction 
in addition to the \(^1S_{0}\) \(nn\) interaction.
At next-to-leading order the interactions gain additional terms and the \(^2S_{1/2}\) \(n\alpha\) interaction is added.
Our treatment of the dynamics of this effective three-body system is based
on the Faddeev formalism.
In addition to the two-body interaction a three-body force is used for
renormalization of the three-body sector. This is all quite similar to 
Ref.~\cite{Li:2023hwe}, but here our focus is on the 
\(E1\) response of \(^6\)He as a case study to highlight the features of
the finite-range EFT. We note that Ref.~\cite{Pinilla:2024xwz} found a rather strong cutoff dependence of the \(E1\) spectrum in Halo EFT, whose origin requires further study.

In impulse approximation the \(E1\) breakup spectrum is directly obtained from the ${}^6$He 
wave function. This calculation can be carried out straightforwardly in the
standard implementation of Halo EFT, as long as one does not attempt to 
evaluate the correction terms stemming from the energy-dependent interaction.
We first calculate the impulse-approximation \(E1\) strength distribution shape in ``standard''
Halo EFT and use it as a cross-check for our finite-range Halo EFT results.

The effect of final-state interactions in the \(E1\) response is included in our calculation using M\o ller operators, as was also done in our earlier work on ${}^{11}$Li~\cite{Gobel:2022pvz}. First we consider the effect on the E1 spectrum that including these interactions through the application of one M\o ller operator to the final-state wave function has. 
We then show how to evaluate products of Møller operators: a more sophisticated approach is needed than in the \(^{11}\)Li case because the \(p\)-wave FSI should be considered at leading order in the \(^6\)He system. Additionally, we consider multiple partial-wave components of the initial state.
A further extension of the FSI calculation in Ref.~\cite{Gobel:2024ovk} is that here we consider multiple interactions in the $nc$ subsystem.
We also compare results from this M\o ller operator product approach (MOPA) to those obtained from truncation of the multiple-scattering series.
For these calculations we developed and implemented a generic scheme that makes use of an expression for the different-spectator overlap.

The paper is structured as follows. In Sec.~\ref{sec:found}, we summarize the formalism used to compute the E1 strength distribution of ${}^6$He, $\frac{d B(E1)}{dE}$. We present the pertinent aspects of Halo EFT in Subsec.~\ref{ssec:contactEFT}, separable Yamaguchi-type interactions in Subsec.~\ref{ssec:FREFT}, the Faddeev equations for the three-body system in Subsec.~\ref{ssec:Fd_formalism},  the computation of  $\frac{d B(E1)}{dE}$ from the bound-state wave function of the two-neutron halo in Subsec.~\ref{ssec:fm_e1_distrib}, and the inclusion of final-state interactions in calculations of that observable via the use of M{\o}ller operators in Subsec.~\ref{ssec:fsi}.
A reader less interested in the details of the implementation can skip Subsecs.~\ref{ssec:Fd_formalism},\ref{ssec:fm_e1_distrib}, \ref{ssec:fsi} and continue with  Sec.~\ref{sec:results}.
We begin Sec.~\ref{sec:results} by giving results for  $\frac{d B(E1)}{dE}$ obtained with the ${}^6$He wave function at LO (Subsec.~\ref{ssec:lo_gs_results}) and NLO (Subsec.~\ref{ssec:nlo_gs_results}) in the EFT but without the inclusion of final-state interactions. Then in Subsec.~\ref{ssec:resultswithFSI} we show the impact of final-state interactions in the $nn$ ${}^1$S$_0$ channel, the $n \alpha$ ${}^2P_{3/2}$ (leading-order effects), and the ${}^2S_{1/2}$ channels (a next-to-leading-order effect). Subsection~\ref{ssec:power_counting} verifies that the size of NLO correction is consistent with the nominal expansion parameter of the EFT, and Subsec.~\ref{ssec:cmp_expm} then compares our final result to the data of Ref.~\cite{Sun:2021efo}.
We summarize and provide an outlook in Sec.~\ref{sec:conclusion}. The first four Appendices describe details of the calculations: the representation of the antisymmetrization operator in the Faddeev equations (App.~\ref{ap:Aij}), the matrix element of the E1 operator in our partial-wave basis (App.~\ref{ap:E1_mel}), the expression for recoupling operations (App.~\ref{ap:recpl}), and the details of calculation of the $nc$ final-state interaction (App.~\ref{ap:nc_fsi}).
Additional FSI results are presented in Apps.~\ref{ap:add_E1_sg_fsis} and \ref{ap:mss}, while the convergence of the expansion in partial waves is analyzed in App.~\ref{ap:pw_trunc}.


\section{Formalism and Methodology}
\label{sec:found}

\subsection{Halo EFT with contact interactions}
\label{ssec:contactEFT}

In the standard realization of Halo EFT, the interactions in the two-body part of the Lagrangian are contact interactions.
A \(s\)-wave interaction between a neutron \(n\) and a core \(c\) in a certain spin channel \(s\) is specified as
\begin{equation}
    \mathcal{L}^{(2)}_{nc} = - C^{(s)}_{nc} \Ke{n c}_{s,m}^\dagger \Ke{n c}_{s,m} \,,
\end{equation}
whereby \(n\) and \(c\) are in general spinor fields and the rectangular brackets describe the coupling to spin \(s\) and projection \(m\)
using Clebsch-Gordan coefficients.
Note that the Einstein summation convention is employed, so that there is a sum over \(m\) and \(s\).
The coupling determining the strength of this interaction is given by \(C_{nc}^{(s)}\).
This interaction Lagrangian is equivalent to an \(nc\) interaction Lagrangian written in terms of the
formation and breakup of a dimer field:
\begin{equation}\label{eq:sw_lag_dimer}
    \widetilde{\mathcal{L}}^{(2)}_{nc} = \Delta_d^{(s)} d_{s,m}^\dagger d_{s,m} - \frac{g_d^{(s)}}{2} \K{ d^\dagger_{s,m} \Ke{n c}_{s,m} + \textrm{H. c.}} \,,
\end{equation}
whereby \(d\) is the dimer field and \(\Delta_d^{(s)}\) a residual mass.
Note that the dimer field is static here. There is no term in the Lagrangian describing the direct motion of the dimer. If a shallow bound state or virtual bound state is present the dimer is then an interpolating
field for such a state. Properties of the state can thus be straightforwardly obtained by computing the dressed dimer propagator.

Interactions in higher partial waves can be realized by adding the corresponding power of the Galilean invariant derivative operator
\(\overleftrightarrow{\partial}_i = \K{\overrightarrow{m} \overleftarrow{\nabla} - \overleftarrow{m} \overrightarrow{\nabla}}_i / \K{\overleftarrow{m}+\overrightarrow{m}}\).
As an example, a \(p\)-wave interaction in a channel of spin \(s\) and overall angular momentum \(j\) is given by
\begin{equation}
    \mathcal{L}^{(2)}_{nc} = - C_{nc}^{(1,s)j} \Ke{n \K{\ci \overleftrightarrow{\partial}} c}_{(1,s)j,m}^\dagger \Ke{n \K{\ci \overleftrightarrow{\partial}} c}_{(1,s)j,m} \,,
\end{equation}
whereby the rectangular brackets with the four indices denotes now the coupling of the spins to the overall spin \(s\), and then of \(s\) to \(l=1\) to form the overall angular momentum \(j\) with its projection \(m\).
In the higher partial waves a formulation in terms of dimer fields in possible, too.

If the \(nc\) system has a low-energy bound state or resonance recasting the Lagrangian in terms of a (spin-1) dimer, \`a la  \cref{eq:sw_lag_dimer} is again useful. 
However, neither the \(p\)-wave contact interaction Lagrangian nor the equivalent dimer formulation has enough couplings to regularize and renormalize the \(nc\) propagator.
They thus cannot represent a consistent leading-order EFT for a system---such as the \(n\alpha\) system---that has a low-energy $p$-wave bound state or resonance. 
For those reasons, we introduce an energy dependence in the Lagrangian. This provides the necessary operators in the dimer propagator to absorb the additional divergence that
appears in the $p$-wave case \cite{Bertulani:2002sz,Bedaque:2003wa}.
In the language of the dimer formalism this means making the dimer dynamic.

The Lagrangian that performs this task reads
\begin{align}
    \widetilde{\mathcal{L}}^{(2)}_{nc} &= d_{(1,s)j,m}^\dagger \K{w_d^{(1,s)j} \K{\ci \partial_0 + \frac{\nabla^2}{2M_{nc}} } + \Delta_d^{(1,s)j} } d_{(1,s)j,m} \nonumber \\
      &\quad - \frac{g_d^{(1,s)j}}{2} \K{ d_{(1,s)j,m}^\dagger \Ke{n \K{\ci \overleftrightarrow{\partial}} c}_{(1,s)j,m} - \textrm{H. c.}} \,,
\end{align}
whereby \(M_{nc}\) is the overall mass of neutron and core.
The dynamics of the \(nc\) system is then determined by resumming all diagrams contributing to the \(nc\) scattering amplitude, which in fact means computing the 
dressed dimer propagator for this case. The (two) UV divergent integrals that appear 
can be regularized using a regularization scale \(\Lambda\) and then renormalized by making the couplings \(w_d^{(1,s)j}\) and \(\Delta_d^{(1,s)j}\) running couplings, i.e., functions of \(\Lambda\).
This \(\Lambda\) dependence of the couplings is determined by extracting the t-matrix from the full two-body propagator and matching the on-shell t-matrix with the expression based on the effective range expansion (ERE) and experimentally determined ERE parameters (See, e.g., the review \cite{Hammer:2017tjm} for details).

\subsection{Yamaguchi interactions}

\label{ssec:FREFT}

An alternative to energy-dependent interactions are finite-range interactions.
Making the interaction finite range increases the number of tunable parameters, and therefore makes it possible---as an energy-dependent interaction also does---
to reproduce two effective-range expansion parameters exactly at leading order.
We use for that purpose the Yamaguchi potential in momentum space, because it is well established.
Its Lagrangian reads
\begin{equation}
    \mathcal{L}^{(2)}_{nc} = - \lambda_{nc}^{(1,s)j} \Ke{n \, g^{(1,s)j}_{nc}{\K{-\ci \overleftrightarrow{\partial}}} c}_{(1,s)j,m}^\dagger \Ke{n \, g^{(1,s)j}_{nc}{\K{\ci \overleftrightarrow{\partial}}} c}_{(1,s)j,m} \,.
\end{equation}
The \(g\) are the form factors and have in the case of the Yamaguchi interaction the following form \cite{Yamaguchi:1954mp,Yamaguchi:1954zz}:
\begin{equation}
    g^{(l,s)j}{\K{p}} \coloneqq p^l \frac{\K{\beta^{(l,s)j}}^4}{ \K{p^2+\K{\beta^{(l,s)j}}^2}^2 } \,.
\end{equation}
To aid the reader in keeping track of the presence of these form factors $g$ we have included them in the Lagrangian, with the understanding that if the Lagrangian is to be employed as a local operator then $p$ must be interpreted as a derivative operator. The Lagrangian could equally well be used without the additional factors $g$ to construct the potential and then the Yamaguchi regularization added afterwards.

In either case the resulting potential's matrix element in quantum mechanical notation reads
\begin{equation}
    \mel{p;\K{l,s}j,m}{V^{\K{\bar{l},\bar{s}}\bar{j}}}{p';\K{l',s'}j',m'} = \kd{l}{l'} \kd{s}{s'} \kd{j}{j'} \kd{m}{m'} 
      \kd{l}{\bar{l}} \kd{s}{\bar{s}} \kd{j}{\bar{j}} \, g^{\K{l,s}j}{\K{p}} \lambda^{\K{l,s}j} g^{\K{l,s}j}{\K{p'}} \,,
\end{equation}
which shows that the potential is separable and of rank one.
Moreover, it is diagonal in the partial waves.

Also in the case of the Yamaguchi potential, we want to determine the potential parameters from the ERE parameters describing the on-shell t-matrix.
Therefore, we have to obtain the corresponding t-matrix, which is given by the Lippmann-Schwinger equation.
It follows that the t-matrix of a rank-one separable potential takes the form
\begin{equation}
    \mel{p;\K{l,s}j,m}{t^{\K{\bar{l},\bar{s}}\bar{j}}}{p';\K{l',s'}j',m'} = \kd{l}{l'} \kd{s}{s'} \kd{j}{j'} \kd{m}{m'} 
        \kd{l}{\bar{l}} \kd{s}{\bar{s}} \kd{j}{\bar{j}} \, g^{\K{l,s}j}{\K{p}} \tau^{\K{l,s}j}{\K{E}} g^{\K{l,s}j}{\K{p'}} \,,
\end{equation}
whereby the energy dependence is captured by the reduced t-matrix \(\tau^{\K{l,s}j}{\K{E}}\).
Moreover, since we carry out the three-body calculations in the momentum-space Faddeev formalism we need expressions for the t-matrices of the interactions anyway.

While the potential is energy-independent, the Lippmann-Schwinger equation gives rise to an energy dependence of the t-matrix.
The reduced t-matrix \(\tau_i{\K{E}}\) captures the energy dependency completely.\footnote{For simplicity, we have replaced the multiple subscripts by a single index.}
It is given by the Lippmann-Schwinger equation for separable potentials:
\begin{equation}
    \tau_i^{-1}{\K{E}} = \lambda_i^{-1} + 8\pi \mu \rint{\q} \frac{\K{g_i{\K{q}}}^2}{q^2 - 2\mu E - \ci \epsilon} \,.
\end{equation}
Note that the energy dependence of the t-matrix arises naturally from the solution of the Lippmann-Schwinger equation.
Thus, in contrast to energy-dependent potentials, energy-dependent t-matrices require no modifications to standard quantum mechanics---as long as they arise from energy-independent potentials.

We reiterate that while we use Yamaguchi interactions here for convenience, other finite-range
interactions characterized by a coupling constant and a range would
work equally well.

\subsection{Three-body systems in Halo EFT}
\label{ssec:Fd_formalism}

After having discussed the two-body sector, contact interactions and Yamaguchi interactions,
we want to give a brief overview of the three-body calculations necessary to describe \(^6\)He on the basis of these interactions
as well as the calculation of the \(E1\) strength distribution.

The first ingredient for solving the three-body system is a basis describing the states.
We use Jacobi momenta specifying the relative momenta in the three-body system.
The basis is introduced with respect to a so-called spectator nucleus \(i\).
It consists of two momenta: the momentum \(\v{p}_i\) among the remaining nuclei \(j\) and \(k\) as well as
the momentum \(\v{q}_i\), the momentum between the spectator \(i\) and the center of mass of \(j\) and \(k\).
The Jacobi momenta are expressed in terms of the momenta \(\v{k}_i\) of the core and the two neutrons as follows:
\begin{align}
    \v{p}_i &\coloneqq \mu_{jk} \K{\frac{\v{k}_j}{m_j} - \frac{\v{k}_k}{m_k}} \,, \\
    \v{q}_i &\coloneqq \mu_{i(jk)} \K{ \frac{\v{k}_i}{m_i} - \frac{\v{k}_j + \v{k}_k}{M_{jk}} }\,,
\end{align}
where \(\mu_{ij} \coloneqq m_i m_j / \K{m_i + m_j}\) and \(\mu_{i(jk)} \coloneqq m_i M_{jk} / \K{m_i + M_{jk}}\).
\(M_{jk}\) is the total mass of particles \(j\) and \(k\).
To get a complete basis, we also need to specify the spin states.
For our calculations, where at certain orders the interactions are only present in certain partial waves, a partial-wave basis is especially useful.
For the three-body system we collect the relevant quantum numbers in multiindices \(\Omega\).
They consist of the following quantum numbers:
\begin{equation}\label{eq:multi}
    \Omega = \Ke{ \K{l, \Ke{s_1, s_2} s}j \K{\lambda, \sigma} I; J,M} \,.
\end{equation}
When applied with some spectator \(i\), i.e., in the form \(\iket{\Omega}{i}\), \(l\) is the relative orbital angular momentum quantum number within the \(jk\) pair.
The spins of the constituents of this pair are \(s_1\) and \(s_2\). They form a total pair spin of \(s\).
In our \(jI\) coupling scheme, the overall spin of this \(jk\) subsystem is well defined and denoted by \(j\).
The relative orbital angular momentum quantum number for the motion of particle \(i\) relative to the center of mass of \(jk\) is given by \(\lambda\).
Together with the spin \(\sigma\) of the spectator, the overall angular momentum of that particle is given by \(I\).
The spins \(j\) and \(I\) form the total angular momentum of the three-body system \(J\).
Its projection is given by \(M\).

Note that the two-body interactions act in a particular partial wave of the corresponding subsystem. 
When computing a particular bound state of the three-body system the total angular momentum quantum numbers \(J\) and \(M\) are also fixed, as is the
quantum number \(\sigma\) for a given spectator. This means that all quantum numbers except for \(\lambda\) and \(I\) are determined.
Depending on the values of \(J\), \(M\), and \(\sigma\) and the constraint by overall parity, the fixed quantum numbers might be already
enough to constrain the allowed values of \(\lambda\) and \(I\) to one value for each of the two.
In this case one can without loss of generality assign a complete multiindex in the three-body system to the interaction channel.
These constants\footnote{
    Note that from this point on in the manuscript the index \(i\) does not denote the spectator, with which the contained quantum numbers are applied, but rather the interaction
    channel to which this constant multiindex belongs to.
    The spectator of this channel \(i\) is given by \(\mathcal{S}{(i)}\).
} are in the following denoted by \(\Omega_i\).
If we use multiindices only for the two-body subsystem, we use the symbol \(\omega\).
It has the same order of quantum numbers as specified in \cref{eq:multi}, but ends with \(j\).
However, even if not all quantum numbers would be constrained, then one could just proceed by assigning the interaction
to multiple three-body partial-wave channels consistent with the constraints.

After having discussed the basis, the second part is obtaining the state or, in practice, representations of it.
For that purpose, we perform a Faddeev treatment, where the full state \(\ket{\Psi}\) is decomposed
into Faddeev components 
\begin{equation}\label{eq:Fi_deq}
    \ket{F_i} \coloneqq \K{G_0 t_i}^{-1} G_0 V_i \ket{\Psi} \,.
\end{equation}
The Faddeev components fulfill the Faddeev equations, which are equivalent to the Schrödinger equation:
\begin{equation}\label{eq:Fd}
    \ket{F_i} = \sum_{j \neq i} G_0 t_j \ket{F_j} \,.
\end{equation}
Often the index takes values such as \(n\) or \(c\) meaning that the different interactions and their respective Faddeev components are denoted by the spectator particle.
However, here at next-to-leading order, we have two \(nc\) interactions: the \(n\alpha\) interaction in \(^2P_{3/2}\) from LO as well as the interaction in \(^2S_{1/2}\), which enters at NLO.
To distinguish the interactions and their Faddeev components while keeping the previous indices, one would need additional indices.
We use an alternative scheme where we just count the interactions and Faddeev components by numbers.
When going into a representation, one  not only needs to index the Faddeev components but also to index the respective spectator.
Since this information is no longer stored in the index itself, we introduce the function \(\mathcal{S}{(i)}\), which returns the spectator for each channel index \(i\).

To illustrate that \cref{tab:int_ch} contains an overview of the two-body interactions used in our calculations.

\begin{table}[H]
    \centering
    \caption{Overview of the two-body interactions in the calculations with focus on the partial-wave structure.}
    \label{tab:int_ch}
    \begin{tabular}{cccccc}
        \toprule
            \(i\) & \(\mathcal{S}{(i)}\) & \(^{2s+1}l_j\) & \(\omega_i\) & \(\Omega_i\) & order \\
        \midrule
            0 & \(c\) & \(^1S_0\)     & \((0, [1/2, 1/2]0)0\)   & \((0, 0)0, (0, 0)0; 0,0 \)         & LO \& NLO \\
            1 & \(n\) & \(^2P_{3/2}\) & \((1, [1/2, 0]1/2)3/2\) & \((1, 1/2)3/2, (1, 1/2)3/2; 0,0 \) & LO \& NLO \\
            2 & \(n\) & \(^2S_{1/2}\) & \((0, [1/2, 0]1/2)1/2\) & \((0, 1/2)1/2, (0, 1/2)1/2; 0,0 \) &       NLO \\
        \bottomrule
    \end{tabular}
\end{table}

Note that for each \(V_i\) or equivalently \(t_i\) with \(\mathcal{S}{(i)} = n\), there is an \(V'_i\) or \(t'_i\) related via antisymmetry:
\(V'_i = \pmo V_i \pmo\) and \(t'_i = \pmo t_i \pmo\) with \(\pmo\) being the \(nn\) permutation operator.
In a system with \(M\) \(nn\) interactions and \(N\) \(nc\) interactions, this would result in \(M+2N\) interaction channels and as many
Faddeev components.
However, from the symmetry relation for the interaction and the antisymmetry of the overall state \(\ket{\Psi}\) follows also one for the components: \(\ket{F'_i} = -\pmo \ket{F_i}\).
One can make use of that explicitly and devise Faddeev equations adjusted for that case, which need only \(M + N\) components.
To account for the antisymmetrization one can introduce the operator \(A_{ij}\) and write the Faddeev equations as
\begin{equation}\label{eq:Fd2}
    \ket{F_i} = \sum_{j} G_0 A_{ij} t_j \ket{F_j} \,.
\end{equation}
A definition of that operator can be found in Appendix \ref{ap:Aij}.

In this context, we want to discuss the finite-range interactions in a bit more detail.
We will use the finite-range Yamaguchi interaction only in these cases where we need to reproduce two effective-range expansion parameters. That is the \(^2P_{3/2}\) \(nc\) interaction at LO and at NLO as well as the \(^1S_0\) \(nn\) interaction at NLO.
In the other cases, the \(^1S_0\) \(nn\) interaction at LO as well as the \(^2S_{1/2}\) \(nc\) interaction entering at NLO, we use momentum-space contact interactions with Heaviside regulators.
The form factor \(g_i{\K{p}}\) in those cases is given by \(g_i{\K{p}} = p^{l_i} \Theta{\K{\Lambda - p}}\) with the regularization scale \(\Lambda\).
These contact interactions can be seen as a special case of the Yamaguchi interaction when one parameter is sufficient.
In future calculations at higher order, where more ERE parameters need to be reproduced, one might employ modified versions of the Yamaguchi interaction with more complicated form factors giving rise to more parameters.
An overview of the ERE parameters reproduced by the different interactions is given in \cref{table:2Binteractions}.
Note that the unitarity term of the \(^2P_{3/2}\) \(nc\) interaction, although being NLO, is not removed at LO, since we would have to include it in our NLO calculation anyway. In contrast, we do not include the unitarity term in the  \(^2S_{1/2}\) \(nc\) $t$-matrix, since that effect is beyond the order we work to here.

\begin{table}[H]
    \centering
    \caption{Overview of the two-body interactions in the calculations with focus on the effective-range expansion (ERE) terms reproduced by the interaction. The unitarity term of the \(^2P_{3/2}\) \(nc\) interaction at LO is in gray color as according to the power counting it is not LO but we included it already at that order in order to have an interaction fulfilling unitarity.}
    \begin{tabular}{ccccc}
        \toprule
            \(i\) & \(\mathcal{S}{(i)}\) & \(^{2s+1}l_j\) & LO ERE terms & NLO ERE terms \\
        \midrule
            0 & \(c\) & \(^1S_0\)     & \( 1/a_{nn}^{(0)} + \ci k\)   & \( 1/a_{nn}^{(0)} + \ci k - \oh r_{nn}^{(0)} k^2 \) \\
            1 & \(n\) & \(^2P_{3/2}\) & \( 1/a_{nc}^{(1; 3/2)} -\frac{1}{2} r_{nc}^{(1; 3/2)} k^2 \textcolor{gray}{+\ci k^3}\)   & \(1/a_{nc}^{(1; 3/2)} - \frac{1}{2} r_{nc}^{(1; 3/2)} k^2 + \ci k^3\) \\
            2 & \(n\) & \(^2S_{1/2}\) &  & \(1/a_{nc}^{(0)}\) \\
        \bottomrule
    \end{tabular}
    \label{table:2Binteractions}
\end{table}

For the actual calculations one needs representations of the states with respect to some basis.
The following representation, the so-called Faddeev amplitude, is commonly used:
\begin{equation}\label{eq:Fiq_def}
    F_i{\K{q}} \coloneqq \rint{\p} g_i{\K{p}} \ibraket{\mathcal{S}{(i)}}{p,q;\Omega_i}{F_i}{} \,.
\end{equation}
In this representation the Faddeev equations \cref{eq:Fd2} read
\begin{align}\label{eq:Fd_repr}
    F_i{\K{q}} = 4\pi \sum_{j} \rint{\qp} X_{ij}{\K{q, q'; E_3}} \tau_{j}{\K{q'; E_3}} F_j{\K{q'}} \,,
\end{align}
whereby \(\tau_j{(q; E_3)}\) is the three-body embedded version of the reduced t-matrix element given by 
\(\tau_j{(q; E_3)} = \tau_j{(E_3-q^2/\K{2\mu_{j(ki)}})}\).
The integral over \(q'\) in this equation is cut off at the three-body cutoff \(\Lambda\).
This scale is also used in the Heaviside regulator of the contact interactions mentioned before.
The \(X_{ij}\) are the so-called exchange kernels, they stem from matrix elements of three-body free Green's function and recoupling between basis states of different spectators:
\begin{equation}
    X_{ij}{\K{q, \qp; E_3}} \coloneqq \rint{\p} g_{i}{\K{p}} \rint{\pp} g_{j}{\K{\pp}} \imel{\mathcal{S}{(i)}}{p,q;\Omega_i}{ G_0{\K{E_3}} A_{ij} }{\pp, \qp;\Omega_j}{\mathcal{S}{(j)}} \,.
\end{equation}

Once the Faddeev amplitudes have been obtained, one can obtain the representations of the full state \(\ket{\Psi}\), the wave functions
\begin{equation}
    \Psi_{\mathcal{S};\Omega}{\K{p,q}} \coloneqq \ibraket{\mathcal{S}}{p,q;\Omega}{\Psi}{}
\end{equation}
using
\begin{equation}
    \ket{\Psi} = \sum_i G_0 B_i t_i \ket{F_i} \,,
\end{equation}
which is based on \cref{eq:Fi_deq}.
The operator \(B_i\) is necessary due to the explicit antisymmetrization approach.
For \(i\) with \(\mathcal{S}{(i)}=n\) it is \(\id - \pmo\), while for channels with \(\mathcal{S}{(i)}=c\) it is the identity operator.
More information on the calculation of the \(^6\)He ground state in Halo EFT in the Faddeev formalism can be found, e.g., in Refs. \cite{Ji:2014wta,Gobel:2019jba,Gobel:2021pvw}.
Note that these sources use slightly different notation as they have only one interaction channel per spectator.
The kernel functions \(X_{ij}\) are evaluated numerically by implementing the analytically simplified form given in chapter 9 of Ref. \cite{Gobel:2024ovk}.
Our Faddeev implementation is suitable for arbitrary many separable interactions of various forms.
We checked that it reproduces the so-called spectator functions, projections of the Faddeev wave function components, that were computed in Ref.~\cite{Ghovanlou:1974zza} (see Fig.~4 and Tab.~VII of that reference).

\subsection{\texorpdfstring{Details regarding the \(^6\)He bound-state calculation}{Details regarding the 6He bound-state calculation}}
\label{ssec:details_bd_st_calc}

For our calculations we use the following parameters.
In the \(nn\) case we use a scattering length of -18.7~fm (see Refs.~\cite{GonzalezTrotter:1999zz,GonzalezTrotter:2006wz} and for an overview, e.g., Ref.~\cite{Gobel:2021pvw}) and an effective range of \(r_{nn} \approx r_{np}{(^1S_0)} = 2.73\)~fm \cite{PrestonBhaduri}.
For the \(^2P_{3/2}\) \(nc\) interaction we use the parameterization of the t-matrix in terms of \(\gamma_1 = -r_1/2\) and \(k_R=\sqrt{2/(a_1 r_1)}\) given in Ref.~\cite{Ji:2014wta}.
The parameters in use are \(a_1 = -62.951\)~fm\(^3\) and \(r_1 = - 0.8819\)~fm\(^{-1}\).
For the \(^2S_{1/2}\) \(nc\) interaction the scattering length is given by 2.4641~fm.
All parameters for the \(n\alpha\) system are taken from Ref.~\cite{Arndt:1973ssf}.

We use a three-body force in order to renormalize the three-body system.
The two-neutron separation energy of 0.975~MeV is employed as input~\cite{Brodeur:2011sam}.
We follow Ref.~\cite{Ji:2014wta} and implement 
the three-body force via the replacement \(X_{11}{\K{q, q'}} \rightarrow X_{11}{\K{q, q'}} + q q' h{(\Lambda)}\), i.e., we modify the diagonal piece of the three-body kernel in the  \(^2P_{3/2}\)-\(n\) channel. Other implementations were explored in Ref.~\cite{Ryberg:2017tpv} and give equivalent results at leading order.

Usually a three-body cutoff of 750~MeV is chosen and convergence checks are based on comparisons with calculations using 500~MeV.
The three-body cutoff is also used as the cutoff for the Heaviside regulators of the contact interactions.

\subsection{\texorpdfstring{\(E1\) strength distribution}{E1 strength distribution}}
\label{ssec:fm_e1_distrib}

We use this finite-range Halo EFT to study the \(E1\) strength distribution \(\dd B(E1) / \dd{E}\) of \(^6\)He.
This strength distribution is the ingredient of the Coulomb dissociation cross section which is specific to a nucleus.
The relation is given by
\begin{equation}
    \frac{\dd \sigma}{\dd E} = \frac{16}{9} \frac{\pi^3}{\hbar c} N_{E1}{\K{E_\gamma}} \frac{\dd B(E1)}{\dd E} \,,
\end{equation}
whereby \(N_{E1}\) is the virtual photon number for the \(E1\) transition and \(E_\gamma\) is the photon energy.
The total kinetic energy of the free three-body system after breakup is given by \(E\). Consequently, the relation \(E = E_\gamma - S_{2n}\) holds.
The \(E1\) strength itself reads (see, e.g., Ref.~\cite{Forssen:2001rj})
\begin{equation}\label{eq:E1_strength_distrib}
    \frac{\dd B(E1)}{\dd{E}} = \frac{1}{2J_i + 1} \sum_{M_i} \sum_{\mu} \int \dd{\tau_f} \left| \mel{f}{\mathcal{M}_{E1,\mu}}{i;J_i,M_i} \right|^2 \delta{\K{E-E_f}} \,,
\end{equation}
where \(f\) is a multiindex containing quantum numbers and kinematic variables to denote all the possible final states.
With \(\int \dd{\tau_f}\) one integrates and sums over these free variables and indices.
The initial state with its overall spin quantum numbers is given by \(\ket{i;J_i,M_i}\).
The operator describing the \(E1\) breakup is denoted by \(\mathcal{M}_{E1,\mu}\).

The \(E1\) operator reads 
\begin{equation}
    \mathcal{M}_{E1,\mu} = e^2 Z_c^2 \v{r}_c \y{1}{\mu}{\v{r}_c} \,,
\end{equation}
with \(e\) being the elementary charge and \(Z_c\) the charge number of the halo's core.

For the initial state, we want to be able to consider multiple partial-wave components.
In the case of initial states with \(\lambda \neq 0\), the corresponding quantum number of the final state \(\lambda'\) can take multiple values and
thereby multiple partial-wave states after the transition are possible.
For these reasons, it useful to have an expression for the \(E1\) operator between arbitrary partial-wave states (in \(jI\) coupling). We obtained
\begin{align}\label{eq:ms_jI_mel_E1_op}
    & \imel{c}{\pqp;\Omega'}{r_c \y{1}{\mu}{r_c}}{\pq;\Omega}{c} \nonumber \\
    &\quad = \kd{s}{s'} \kd{\sigma}{\sigma'} \kd{l}{l'} \kd{j}{j'} \ci^{\lambda'-\lambda}  \frac{\delta{\K{p'-p}}}{p'^2} D_{q'}^{\K{\lambda, \pm 1} } \frac{\delta{\K{q'-q}}}{q'^2} \sqrt{\frac{3}{4\pi}} f_c \sqrt{\hat{I} \hat{I'}} \sqrt{\hat{\lambda} \hat{J}}
    \cgcc{\lambda,0,1,0}{\lambda',0} \cgcc{J,M,1,\mu}{J',M'} \nonumber \\
    &\qquad \times \K{-1}^{2s + \sigma + \lambda' + j'} \K{-1}^{-J'} 
      \begin{Bmatrix} 1 & I' & I \\ j' & J & J'\end{Bmatrix} \begin{Bmatrix} 1 & I' & I \\ \sigma & \lambda & \lambda' \end{Bmatrix} 
      \,.
\end{align}
where the curly brackets denote Wigner \(6j\) symbols and \(f_c = 2 / \K{A+2} \) is a conversion factor.
The hat over a quantum number has the following meaning: \(\hat{l} = 2l+1\).
Note that we made use of the differential operator
\begin{equation}\label{eq:def_dop}
    D_{q'}^{\K{\lambda, d} } \coloneqq - d {q'}^{d \lambda-(1-d)/2} \partial_{q'} q'^{-d\lambda+(1-d)/2}
\end{equation}
with \(d\) being either \(-1\) or \(+1\).
Additional details on the derivation of \cref{eq:ms_jI_mel_E1_op} and some intermediate results can be found in Appendix \ref{ap:E1_mel}.

It is important to note that the final \(E1\) strength distribution has to fulfill the so-called
non-energy weighted sum rule (see, e.g., Ref.~\cite{Forssen:2001rj}).
For the formulation of that it is useful to introduce the cumulative \(E1\) distribution given
by 
\begin{equation}
    B{(E1)}{(E)} \coloneqq \int_0^{E} \dd{E'} \frac{\dd B{(E1)}}{\dd E'} \,.
\end{equation}
The sum rule then reads
\begin{equation}\label{eq:e1_sum_rule}
    \lim_{E \to \infty} B{(E1)}{(E)} = \frac{3}{4\pi} Z_c^2 e^2 \expval{r_c^2} \,.
\end{equation}
It can be derived by applying first the energy integral to the definition of the \(E1\) strength distribution and then using the completeness of the final states. Finally, the close relation between the \(E1\) operator and \(\vec{r}_c\) is used.
The sum rule provides an important consistency check for the results.

\subsection{\texorpdfstring{\(\alpha\) distance and charge radius of \(^6\)He}{alpha distance and charge radius of 6He}}

In order to obtain the radius we calculate the form factor and extract it from the form factor's derivative in \(k^2\).
The form factor is given by
\begin{equation}
    \mathcal{F}_i{\K{\v{k}}} \coloneqq 
    \expval{P_{\v{k}}^{(i)}} = \mel{\Psi}{ \K{ \inttd{p} \inttd{q}  \iketbra{\vpqk}{i}{\vpq} \otimes \spid } }{\Psi} \,.
\end{equation}
Hereby, \(P_{\v{k}}^{(i)}\) denotes the operator that shifts the \(\v{q}\) momentum by \(\v{k}\).
Next, we define the angular average of the form factor:
\begin{equation}
    \widetilde{\mathcal{F}}_i{\K{k^2}} = \frac{1}{4\pi} \angint{k} \mathcal{F}_i{\K{\v{k}}} \,.
\end{equation}
The root-mean-square distance between the spectator and the remaining particles can then be
extracted via
\begin{equation}
 \expval{y_i^2} = \K{-6} \partial_{k^2} \widetilde{\mathcal{F}}_i{\K{k^2}} \big|_{k^2=0} \,.
\end{equation}
The relation between \(r_c\) and \(y_c\) is then given by \(r_c = \frac{2}{A+2} y_c\).
The expressions shown here for the angle-averaged form factor are in practice evaluated
in a partial-wave basis, whereby a truncation of the included basis states is applied.

With our calculations we obtain for \(r_c\) at LO the value 1.24~fm (\(l_{\mathrm{max}}=3\); \(l_{\mathrm{max}}=0\) yields 1.29~fm) and at NLO 1.14~fm (\(l_{\mathrm{max}}=3\); \(l_{\mathrm{max}}=0\) yields 1.19~fm).
Depending on the interactions and the method in use Zhukov \textit{et al.} obtained between 1.18~fm and 1.29~fm for \(r_c\) \cite{Zhukov:1993aw}. 
For typical interactions Grigorenko \textit{et al.} obtained values between 1.15~fm and 1.17~fm \cite{Grigorenko:2020jts}.
We are in good agreement with those results.

Experimental data exist for the charge radius of \(^6\)He. Krauth {\it et al.} give  \(  r_{\textrm{ch}}{\K{{}^6\textrm{He}}}=2.0571 \pm 0.0075\)~fm, based on the previously measured isotope shift and their muonic ion measurement of the charge radius of 
\(^4\)He~\cite{Krauth:2021foz}.
Therefore we use the formula
\begin{equation}
    r_{\textrm{ch}}^2{\K{{}^6\textrm{He}}} = r_{\textrm{ch}}^2{\K{{}^4\textrm{He}}} + r_{\textrm{ch}}^2{\K{n}} + r_c^2
\end{equation}
to calculate the charge radius of \(^6\)He from the charge radius of \(^4\)He, the squared charge radius of the neutron and the value for \(r_c\) we have obtained.
The squared charge radius of the neutron is negative and given by \(r_{\textrm{ch}}^2{\K{n}} = - 0.1155 \pm 0.0017\)~fm\(^2\) \cite{ParticleDataGroup:2024cfk}, while 
Ref.~\cite{Krauth:2021foz} measured a \(^4\)He charge radius of \(1.67824 \pm 0.00095\)~fm.
With these values we obtain a LO \(^6\)He charge radius of 2.06~fm and an NLO radius of 2.00~fm. Based on the EFT expansion parameter (see Sec.~\ref{ssec:power_counting}) we assign a LO EFT uncertainty is 0.35 fm, this is reduced to 0.09 fm at NLO. The size of the NLO shift is thus consistent with the power counting, and both the LO and NLO results overlap the highly precise experimental number.

\subsection{Final-state interactions}
\label{ssec:fsi}

As can be seen from \cref{eq:E1_strength_distrib}, besides the matrix elements of the \(E1\) operator, the strength distribution
has two important ingredients.
These are the initial and the final state, respectively their representations in terms of wave functions (in a partial-wave basis).
How the initial-state wave function can be obtained was already discussed in \cref{ssec:Fd_formalism}.
If we assume no interaction between the three free particles, the two neutrons and the \(\alpha\) core, after the \(E1\) breakup,
then the final state is just a (partial-wave projected) plane-wave state, i.e., \(\iket{p,q;\Omega}{c}\).
Interactions following the \(E1\) breakup, the so-called final-state interactions (FSIs), distort the final state.
To take all interactions fully into account, the final state with \(p\) and \(q\) in partial wave \(\Omega\) as boundary condition of that three-body scattering state would be
\begin{equation}
 \moop_V \iket{p,q;\Omega}{c} \,,
\end{equation}
whereby \(\moop_V\) is the M{\o}ller operator based on all the possible FSIs. This means \(V\) is the sum of all potentials.
A more explicit expression for the M{\o}ller operator reads
\begin{equation}
  \moop_V = \id + \lim_{\epsilon \to 0^+} \rint{\p} \rint{\q} G{\K{E_{p,q} + \ci \epsilon}} V \ketbra{p,q}{p,q} \,.
\end{equation}
Note that this \(V\) is also employed by the full Green's function \(G\).

Instead of calculating the full distorted state, we calculate different approximations.
One approach is to evaluate only the effect of one of the FSIs.
This can, for example, be  the \(nn\) or one of the \(nc\) interactions.
Thereby, one can compare the effect of different FSIs.
Another approach is to expand the action of the full M{\o}ller operator on the boundary condition state in terms of the t-matrices and truncate this multiple-scattering series at a certain order.
This expression could then be evaluated in terms of the single summands.
Expanding in the t-matrices is possible because the M{\o}ller operator can  also be  written in terms of the t-matrices instead of the potentials.
However, this expansion is not necessarily unitary.

An alternative proposed in Ref. \cite{Gobel:2022pvz} is to use products of the M{\o}ller operators from single FSIs.
By construction, this is unitary as long as the single M{\o}ller operators are unitary.
We will test all three approximation strategies:
\begin{itemize}
    \item inclusion of single FSIs only (via the corresponding M{\o}ller operator),
    \item direct truncations of the multiple-scattering series (MSS),
    \item approximations based on products of M{\o}ller operators (M{\o}ller-operator product approximation, MOPA).
\end{itemize}

The MOPA approach was already used in Ref. \cite{Gobel:2022pvz} for obtaining the \(E1\) strength distribution of the two-neutron halo \(^{11}\)Li.
In that case the dominant \(nc\) interaction is in the \(s\)-wave.
Therefore there were no issues with the energy dependence.
One might wonder if the expressions for the FSI derived there can be directly used here.
This is not the case as the partial-wave structure of the final-state interactions is different and here we also want to consider multiple initial partial-wave states, while for \(^{11}\)Li we considered one initial partial-wave state.

As a consequence, we will work in \(jI\) coupling also for the FSI part and have the additional complexity that when going from a \(c\)-spectator state
to a \(n\)-spectator state for the \(nc\) interaction \(\lambda\) and \(I\) are not fixed and we have to handle different three-body states
with the same state in the two-body subsystems.

None of these complications affect the evaluation when only the \(nn\) FSI is present, as then we do not need to work out any basis recouplings.
The expression for the non-trivial part of that M\o ller operator is given by
\begin{align}   \label{eq:nnFSI}
    &\imel{c}{p,q;\Omega}{ \K{\Omega_{nn}-\id}^\dagger \mathcal{M}_{E1,\mu}}{\Psi}{}
    = \kd{\K{\Omega}_{nn}}{\omega_{nn}} \sum_{\Omega_i} f_{E1,\mu}^{\K{\Omega,\Omega_i}} g_c{\K{p}} \tau_{nn}{\K{E_p}} \nonumber \\
    &\quad \times \rint{\pt} g_c{\K{\pt}} \frac{1}{E_p - \frac{\pt[2]}{2\mu_{nn}} + \ci \epsilon } D_q^{(\Omega,\Omega_i)} \Psi_{c,\Omega_i}{\K{\pt,q}} \,.
\end{align}
The expression \(\delta_{(\Omega)_{nn}, \omega_{nn}}\) means that only those partial-wave states \(\Omega\) contribute where the two-body part \( (\Omega)_{nn} \) equals the the two-body multiindex of the \(nn\) interaction given by \(\omega_{nn}\).
The pole is treated by using the Sokhotski-Plemelj theorem and a numerical implementation of the principal value integral.

When dealing with any other FSI or combination of FSIs we employ the approach of Ref. \cite{Gobel:2024ovk} (chapter 9.4), where a general formula for expressions involving different-spectator overlaps of the form
\begin{equation}\label{eq:O_op}
    \tof{\mathcal{S}, \mathcal{S}'}{p, q}{\pp,\qp}{\Omega, \Omega'} f{\K{\pp,\qp}} \coloneqq \rint{\pp} \rint{\qp} \ibraket{\mathcal{S}}{p,q;\Omega}{\pp,\qp;\Omega'}{\mathcal{S'}} f{\K{\pp,\qp}}
\end{equation}
was obtained.
As can be seen by looking at the defining expression on the right, the \(|p', q'\) in the superscript of \(\mathcal{T}\) just defines variables
over which one integrates: these are variables which are consumed by the transformation to the new spectator basis.
Applying \(\mathcal{T}\) on \(f\) therefore yields an expression depending on \(p\) and \(q\) but not depending on \(p'\) and \(q'\).

The impact of this transition between spectator bases on the function $f$ can be evaluated by decoupling the states into pure spatial (radial and angular) and pure spin parts.
The different-spectator spin overlap can be then evaluated more directly, while for the spatial part one can insert plane-wave identities for each spectator.
After some simplification we obtain a final result that is a sum over many quantum numbers in which each term is a product of Wigner-6\(j\) and Wigner-9\(j\) symbols and partial-wave projections of a function related to the argument \(f\) of the expression above.
Each partial-wave projection requires only the evaluation of a single integral, thereby greatly reducing the numerical effort over a naive implementation of the spectator-basis transition. 
For self-containedness and to clarify a few typos of Ref.~\cite{Gobel:2024ovk}, the final expression for \( \tof{\mathcal{S}, \mathcal{S}'}{p, q}{\pp,\qp}{\Omega, \Omega'} f{\K{\pp,\qp}} \) is given in Appendix~\ref{ap:recpl}.

Two of the FSIs we are interested in evaluating are the \(nc\) LO \(^2P_{3/2}\) FSI and the NLO \(^2S_{1/2}\) one.
In both cases, the non-identity part of that M{\o}ller operator is given by
\begin{align}\label{eq:nc_fsi_mel}
    &\imel{c}{p,q;\Omega}{ \K{\moop_{nc} - \id}^\dagger \mathcal{M}_{E1,\mu} }{\Psi}{}
    = \sum_{\Omega_i,\Omega_m}   \sum_{\substack{ \Omega' \textrm{ with}\\ \K{\Omega'}_{nc}=\omega_{nc}}} \tof{cn}{p,q}{p',q'}{\Omega, \Omega'}
    g_{n}{\K{p'}} \tau_{nc}{\K{E_{p'}}} \nonumber \\
    &\quad \times \rint{\pt} g_{n}{\K{\pt}} \frac{1}{E_{p'}- \pt[2]/\K{2\mu_{nc}} + \ci \epsilon} 
    \tof{nc}{\pt,q'}{\ppp,\qpp}{\Omega',\Omega_m} D_{q''}^{\K{\Omega_m,\Omega_i}} \Psi_{c,\Omega_i}{\K{\ppp,\qpp}} f_{E1,\mu}^{\K{\Omega_m,\Omega_i}}
    \, .
\end{align}
The expression \(\K{\Omega'}_{nc}=\omega_{nc} \) under the summation sign means that we sum only over those \(\Omega'\) where the \(nc\) part of the multiindex equals the two-body multiindex of the \(nc\) interaction \(\omega_{nc}\).
For the \(^2P_{3/2}\) \(nc\) FSI \(\omega_{nc}\) reads \((1,1/2)3/2\), while for the \(^2S_{1/2}\) FSI it is given by \((0,1/2)1/2\).
If one reads the equation from the lower right to the upper left, the physical process unfolds chronologically into the following steps:
the breakup of the initial three-body system by the \(E1\) operator seen from the core as spectator,
the transition to a neutron as spectator,
the evaluation of the \(nc\) FSI contribution,
and the recoupling to the core as spectator.
The sum over the multiindex \(\Omega_i\) accounts for the different possible initial partial-wave states.
The sums over the multiindices \(\Omega_m\) and \(\Omega'\) reflect that (in general) multiple partial-wave states are possible after the \(E1\) breakup and after the recoupling to a \(n\)-spectator for evaluation of \(nc\) FSI.
Note that the integral over the two-body free Green's function and the transformed wave function contains a singularity at \(\pt = \pp\), i.e., a moving singularity.
Connected by the final transformation from \(n\) spectator back to the \(c\) spectator, it moves with the final-state momenta.
This integral is evaluated by using the Sokhotski-Plemelj theorem yielding an imaginary term and a real term involving a principal value integral
of the free Green's function and the transformed wave function.
The principal value integral is evaluated numerically.
While we choose this notation to give a more direct and transparent glimpse on the computational techniques employed, 
Appendix~\ref{ap:nc_fsi} contains this equation in plain bra-ket notation. (The expression shown here is actually derived from the bra-ket notation expression
by applying the definition in \cref{eq:O_op}.) Equations (\ref{eq:nc_fsi_mel}) and (\ref{eq:nnFSI}) provide us with the expressions needed to evaluate the effect of the single two-body M{\o}ller operators on the E1 matrix element evalaution. 

But the beauty of the recoupling technique developed for evalaution of the \(nc\) FSI is that it inductively extends to the evaluation of products of two or more M{\o}ller operators. 
For that purpose in the product of single M{\o}ller operators each operator is decomposed into its trivial and non-trivial part according to \(\Omega_{ij}^\dagger \eqqcolon \id + \bar{\Omega}_{ij}^\dagger\).
The product is multiplied out yielding a sum of products of the non-trivial parts \(\bar{\Omega}_{ij}^\dagger\).
If written on paper, each summand  corresponds to expressions similar to \cref{eq:nc_fsi_mel} but with more \(\mathcal{T}\) operators, more sums over intermediate states and more integrals with singularities. Theoretically, there is no limit on the number of interactions in the FSI sequence. The only limits are computation time and memory footprint. Any product of non-trivial parts of M{\o}ller operators contains a sequence of recouplings and a sequence of nested integrals with moving singularities.
These are evaluated as described in the previous paragraph. 
Recoupling between different spectators is made whenever necessary using the expression for the \(\mathcal{T}\) operation.
For the intermediate states, all possible partial-wave states are considered.
The computer program automatically does the calculations for a given sequence of FSIs by constructing the integrands and performing the nested integrations and recouplings\footnote{
In principle this approach could be carried out without splitting the M{\o}ller operator into a trivial and non-trivial part. However, the splitting allows us to exploit the FSI properties and truncate the partial-wave bases in the intermediate steps.
}.

This strategy is also applied to the evaluation of the truncated multiple-scattering series, by applying it to each term of the series, which is a product of \(\bar{\Omega}_{ij}^\dagger\), i.e., of  
non-trivial part of M{\o}ller operators.
These terms are then evaluated individually as described in the previous paragraph. 

The specific formula to obtain the overall \(E1\) strength distribution with or without FSI based on the contributions in different final partial waves reads
\begin{equation}\label{eq:e1_distrib_final}
    \frac{\dd B(E1)}{\dd{E}} = \rint{\q} \rint{\p} \frac{1}{2 J_i+1} \sum_{\mu} \sum_{\Omega_f} \left| \sum_{\Omega_i} \imel{c}{p,q;\Omega_f}{\widetilde{\Omega} \mathcal{M}_{E1,\mu} P_{\Omega_i}}{\Psi}{} \right|^2 \delta{\K{E- \frac{p^2}{2\mu_{nn}} - \frac{q^2}{2\mu_{c(nn)}}}} \,,
\end{equation}
where \(\widetilde{\Omega}\) is the approximation to the full-three-body FSI operator.
If no FSI is included, it is the identity operator \(\id\).
The strategy for the computation of the FSI explained above provides a way to evaluate the \(E1\) matrix element
\(\imel{c}{p,q;\Omega_f}{ \tilde{\Omega} \mathcal{M}_{E1,\mu} P_{\Omega_i} }{\Psi}{}\).
We perform a truncation in the included partial-wave states for the initial state (multiindex \(\Omega_i\)) and for the final state (multiindex \(\Omega_f\)).
The condition for the initial states is \(l_i \leq \lmi \, \land \, \lambda_i \leq \lmi\), whereby \(\lmi\) is the truncation parameter.
Similarily, the final states are truncated based on \(\lmf\).
Note that the sum over \(\Omega_f\) is outside the absolute value square, while the one over \(\Omega_i\) (as well as those over intermediate states not written out explicitly) are under the absolute value square.
This is just a consequence of \cref{eq:E1_strength_distrib}.
More information on the truncation can be found in \cref{ap:pw_trunc}.


\section{\texorpdfstring{\(E1\) strength distribution}{E1 strength distribution}}

\label{sec:results}

\subsection{LO ground-state results}

\label{ssec:lo_gs_results}

\begin{figure}[H]
    \centering
    \includegraphics[width=0.92\textwidth]{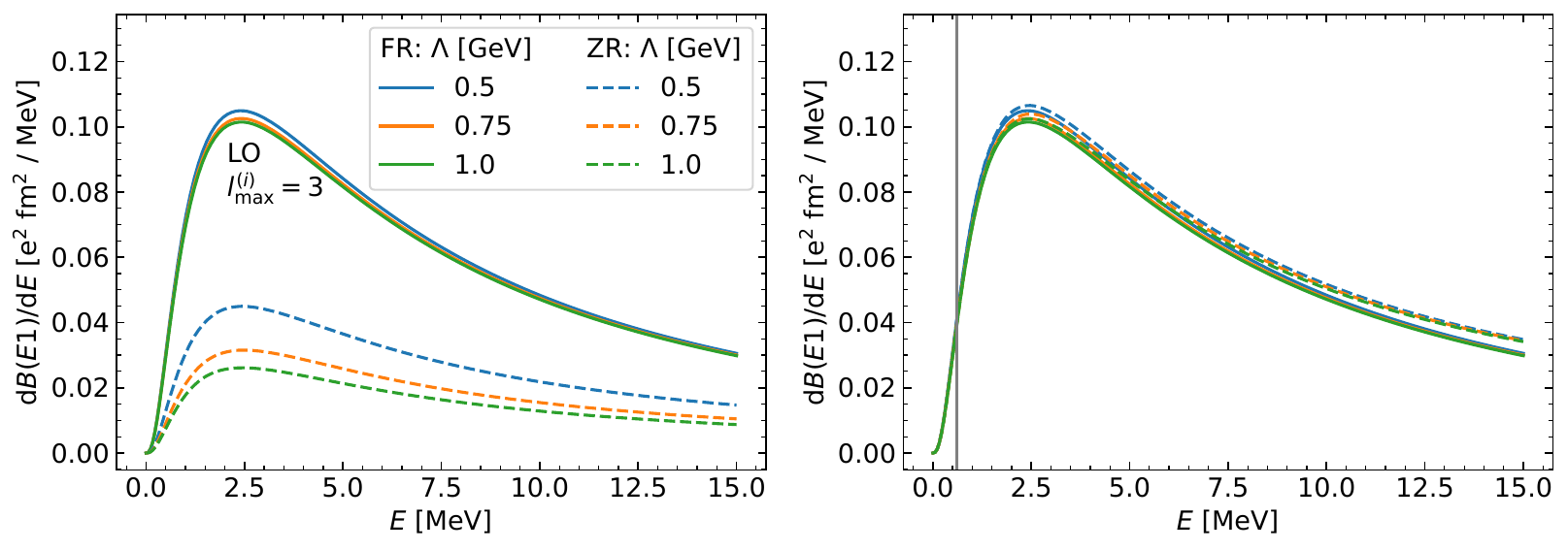}
    \caption{\textit{Left panel:} Leading-order results for the \(E1\) strength distribution obtained in the zero-range (ZR) and finite-range (FR) approaches are shown for different three-body cutoffs \(\Lambda\) between 500~MeV and 1000~MeV. All results are without FSI.
    \textit{Right panel:} Results from the same ZR and FR treatments of two-body interactions and three-body cutoffs.
    The only difference is that,  at the position indicated
    by the vertical line, the ZR results are rescaled to match the values of FR results with the same cutoff.}
    \label{fig:LOFREFT}
\end{figure}

In the left panel of Fig.~\ref{fig:LOFREFT} we show the results for the $E1$ strength distribution of ${}^6$He obtained with our leading-order calculation of its structure in finite-range EFT. These results do not include final-state interactions. The different curves show the results for three-body cutoffs of 500, 750, and 1000 MeV.
Moreover, we show also the results based on the standard zero-range approach to Halo EFT for the same cutoffs.

While the results from the finite-range EFT approach show very little cutoff dependence, the ones from the zero-range approach show a considerable dependence.
This is because the correction terms (see Ref.~\cite{formanek04,Gobel:2019jba}) due to the energy-dependent \(^2P_{3/2}\) interaction are not included.
While their influence on the underlying wave function is small for small momenta, they have a significant influence at high momenta~\cite{Gobel:2019jba}.
If they are ignored they cause the non-convergence of the normalization factor in the cutoff and this is exactly the effect we see here.
As at small energies of the distribution only the wave function at small momenta is relevant, the corrections terms are only missing in a significant way in the normalization factor and the shape should be approximately stable.
This is what we observe in the right panel of Fig.~\ref{fig:LOFREFT}.
There the zero-range results are rescaled to match the finite-range EFT results at the position indicated by the vertical line.
One can see that the shapes approximately agree.
Starting at the matching position the relative deviations increase to about 15\% around \(E=15\)~MeV.
That the deviations in the shape increase with the energy is not surprising as the correction terms which also can modify the shape become more important at higher momenta in the wave function, i.e., higher energies in the distribution.
Interestingly, the influence seems to increase also with the partial waves.
If one truncates at \(\lmi=0\), the deviation at \(E=15\)~MeV is about 6\%.
For \(\lmi=3\), it is about 15\%  at the same energy.

\subsection{NLO ground-state results}
\label{ssec:nlo_gs_results}

At next-to-leading order three additional effects enter the calculation of the ${}^6$He bound state:
\begin{enumerate}
\item The unitarity piece of the $n \alpha$ ${}^2P_{3/2}$ amplitude;

\item The effective range term of the $nn$ ${}^1S_0$ amplitude;

\item The $n \alpha$ ${}^2S_{1/2}$ amplitude.
\end{enumerate}

Of these the first is already included in the FREFT calculation shown in Fig.~\ref{fig:LOFREFT}, since the ${}^2P_{3/2}$ amplitude appearing in the Faddeev equations that are solved for the LO wave function is the full amplitude obtained with the finite-range potential, i.e., it includes the nominally NLO unitarity term, see Table~\ref{table:2Binteractions}. 
This is done in order to have a unitary interaction already at LO.
In general, including higher-order effects in an EFT calculation at a lower order should do no harm, as this effect should be within the uncertainties of the current calculation anyway.

\begin{figure}[H]
    \centering
    \includegraphics[width=0.45\textwidth]{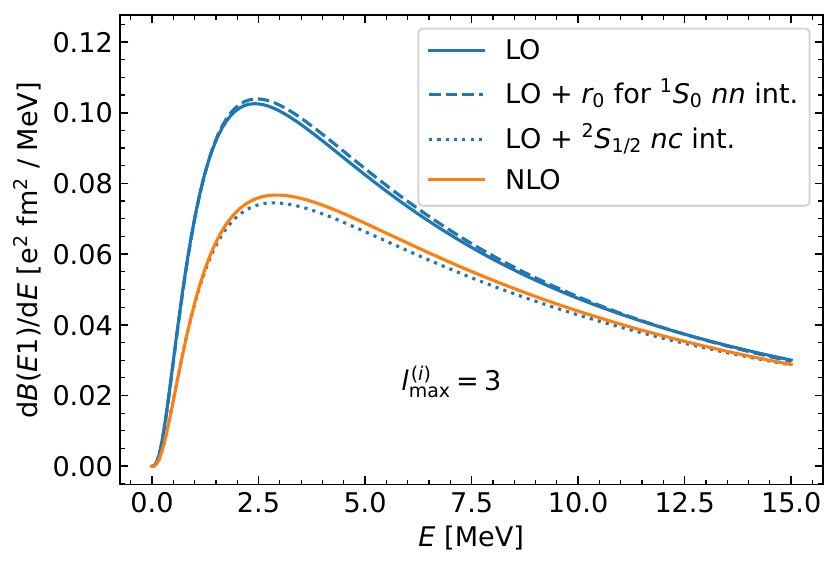}
    \caption{Different \(E1\) strength distribution from the finite-range Halo EFT are shown. All distributions are without FSI. The LO and the NLO result are shown. Moreover, curves stemming from different partial NLO calculations are included to show the effect of different NLO aspects.}
    \label{fig:lo_nlo_gs_cmp}
\end{figure}

We include the effective range in the $s$-wave $nn$ amplitude as described in Subsec.~\ref{ssec:FREFT}: we replace the zero-range $nn$ potential used in the LO FREFT calculation of the previous subsection with a finite-range Yamaguchi potential in which $\beta^{(0,0)0}$ is adjusted to reproduce the physical $nn$ effective range (again, see Table~\ref{table:2Binteractions}). We then employ the resulting amplitude in the three-body calculation. The three-body force is adjusted to keep the two-neutron separation energy of ${}^6$He at its experimental value. This produces the dashed blue curve for the $E1$ strength distribution (as yet without FSI) shown in Fig.~\ref{fig:lo_nlo_gs_cmp}. The leading-order result is also shown there (in blue) for comparison.
The shift due to introduction of a finite ${}^1S_0$ $nn$ effective range in the computation of the initial-state ${}^6$He wave function is below 4\%, 
suggesting that the $nn$ effective range is indeed an NLO effect in this observable. 

Last we consider the incorporation of the \({}^2S_{1/2}\) \(nc\) amplitude in the NLO calculation. We employ
a constant ${}^2S_{1/2}$ amplitude 
within the Faddeev formalism, i.e., the higher-order corrections in this two-body amplitude that make it unitary are not included when computing the NLO correction to the $^6$He ground state, cf. our treatment of the unitarity piece of the \({}^2P_{3/2}\) \(nc\) amplitude. We also again readjust the three-body force so that the binding energy of ${}^6$He remains at the physical value. This produces a result that is independent of the three-body cutoff. 
The NLO effect of the \({}^2S_{1/2}\) \(nc\) channel reduces the  E1 strength 
by about 20\% to 30\% in the peak region,  see the blue dotted curve in Fig.~\ref{fig:lo_nlo_gs_cmp}. While this is the largest NLO effect on the $E1$ strength of the initial state it is still within the expected $\approx 30$\% of an NLO effect.

The convergence of this no-FSI result with the partial waves included in the initial-state wave function is reasonable.
We truncate the partial-wave states based on the condition \(l \leq l_{\mathrm{max}}^{(i)} \, \land \, \lambda \leq l_{\mathrm{max}}^{(i)}\). Going from \(\lmi=1\) to \(\lmi=2\) changes the distribution up to \(E=\)15~MeV by less than 2.5\%. Switching from \(\lmi=2\) to \(\lmi=3\) results in changes smaller than 1.5\%.
As the label inside the figure shows, the curves were produced with the setting \(\lmi=3\).
A more detailed discussion of the truncation can be found in Appendix~\ref{ap:pw_trunc}.

\subsection{Results with FSI}
\label{ssec:resultswithFSI}

\begin{figure}[H]
    \centering
    \includegraphics[width=0.92\textwidth]{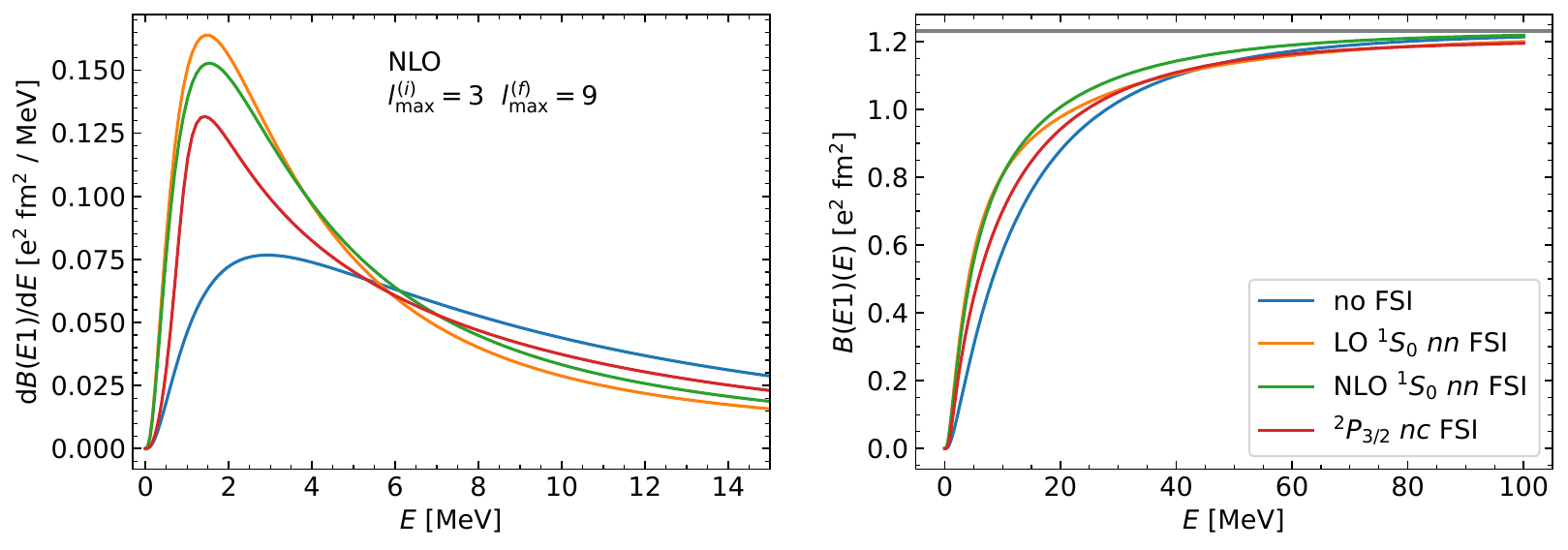}
    \caption{\label{fig:nlo_fsi_cmp1}
      \textit{Left panel:} The blue curve shows the NLO FREFT \(E1\) distribution without FSI. The orange and green curves then show the result when $nn$ FSI is included in its LO (zero range) and NLO (finite range) form. The red curve is the result when the ${}^2P_{3/2}$ $nc$ FSI is included. 
    \textit{Right panel:} The corresponding cumulative \(E1\) distributions, with the gray line showing the expected value for \(B{(E1)}{(E \to \infty)}\) obtained using the \(\expval{r_c^2}\) given by the form factor. Note that all curves correspond to a result that is consistent with the sum rule.}
\end{figure}

We include final-state interactions in the calculation using M{\o}ller operators, see Sec.~\ref{ssec:fsi}. The two channels with LO-size interactions are the $nn$ ${}^1S_0$ and the $nc$ ${}^2P_{3/2}$. The results of applying M{\o}ller operators corresponding to each of these to the NLO \(E1\) amplitude of the previous subsection are shown in Fig.~\ref{fig:nlo_fsi_cmp1}. All curves in Fig.~\ref{fig:nlo_fsi_cmp1} use the ground-state Halo EFT momentum distribution computed at NLO in Halo EFT: they differ only in which FSI is considered.

For energies of the outgoing three-body subsystem that are less than 6 MeV the $nc$ $p$-wave interaction has a  smaller impact on the $E1$ strength distribution than the $nn$ interaction does.
However, its effect is still quite strong and the peak occurs in a similar location for both FSIs.
The LO \(nn\) FSI increases the peak height by about 114\% compared to the curve without FSI, while
the \(p\)-wave \(nc\) FSI causes an increase over the non-FSI result by 72\%.
This is quite different than the situation in the case of the \(E1\) distribution of \(^{11}\)Li, where the \(nc\) FSI, which is in the \(s\)-wave there, is markedly weaker than the \(nn\) FSI.
The finding of the strong \(nc\) FSI in \(^6\)He is in agreement with Kikuchi~\textit{et al.}~\cite{Kikuchi:2010zzb}, where also sizable \(nc\) FSI effects in the strength distribution have been observed.

In addition to the LO version of the \(nn\) FSI also its NLO version is shown. 
According to our FREFT scheme, the LO version parameterized only by the scattering length is given by a contact interaction with Heaviside regulators.
In contrast, the NLO version, which is additionally parameterized by the effective range, is given by a Yamaguchi interaction with a finite $\beta^{(0,0)0}$. 
Changing the FSI from the NLO to the LO form reduces the height of the peak of the E1 distribution by approximately 7\%. There is a compensating increase in the \(E1\) strength at higher outgoing energies. The size of this effect is  again consistent with the identification of $r_{nn} \neq 0$ as an NLO effect in this problem.

Meanwhile the right panel compares the cumulative $E1$ strength with the result for this quantity that is expected from the non-energy-weighted sum rule (see Eq.~(\ref{eq:e1_sum_rule}) in Sec.~\ref{ssec:fm_e1_distrib}). All the curves, without FSI, with LO \(^1S_0\) \(nn\) FSI, with NLO \(^1S_0\) \(nn\) FSI, and with \(^2P_{3/2}\) \(nc\) FSI, nicely fulfill the sum rule.

In the case of the NLO \(^1S_0\) \(nn\) FSI it is also interesting to look at the difference between the here shown finite-range implementation and the zero-range implementation.
Since in the case of the zero-range implementation we are not implementing the correction terms due to the energy dependence, the FSI calculation using it does not conserve probability and we see a significant violation of the sum rule.
Similar observations can be made for the finite-range and the zero-range version of the \(^2P_{3/2}\) \(nc\) FSI.
More details and a corresponding plot (\cref{fig:nlo_fsi_cmp1b}) can be found in Appendix~\ref{ap:add_E1_sg_fsis}.

So far we have focused on the comparison of the no FSI, the \(nn\) FSI, and the \(^2P_{3/2}\) \(nc\) FSI. However, there is also the \(^2S_{1/2}\) \(nc\) FSI.
If one calculates FSI based on the strict NLO version of this interaction, which is missing at that order a unitarity term, one faces problems with non-conservation of probability and a corresponding violation of the sum rule.
Therefore we use already at NLO the interaction including the unitarity term.
However, it turns out that there is still a small violation of the sum rule by about 4\%.
This is due to the fact that this zero-range interaction with its positive scattering length supports a bound state.
The corresponding M{\o}ller operator therefore is isometric but not unitary, because the Hilbert space \(\mathcal{H}\) of eigenstates of the corresponding Hamilton operator \(H\) is not equal to the space of scattering states of this Hamilton \(\mathcal{S}\).
Instead it is the direct sum \(\mathcal{S} \oplus \mathcal{B}\) with \(\mathcal{B}\) being the Hilbert space of bound states of \(H\).
For more information on the difference between isometry and unitarity see, e.g., Ref.~\cite{taylor_st}.
The calculations show that in the region up to 5~MeV the \(^2S_{1/2}\) \(nc\) FSI's effect is even much weaker than the one of the \(^2P_{3/2}\) \(nc\) interaction.
In contrast to the LO FSIs considered so far it almost does not change the peak position of the ground-state distribution.
Another distinction is that it does not enhance the peak but slightly suppresses it.
The corresponding plot is included in Appendix~\ref{ap:add_E1_sg_fsis} (\cref{fig:nlo_fsi_cmp2}).

Finally, we present results based on the M{\o}ller-operator-product approximation (MOPA) approach discussed in Sec.~\ref{ssec:fsi}.
In this approach the full three-body scattering state is approximated by acting with a product of M{\o}ller operators 
corresponding to single FSIs on a (partial-wave projected) plane-wave state.
It has the advantage that it is by construction unitary.
Hereby, we limit ourselves to the FSIs which appeared in the previous plots to be the strongest. These are the \(^2P_{3/2}\) \(nc\) FSI and the \(^1S_0\) \(nn\) FSI. We note that this immediately means 
that the MOPA approach cannot be extended beyond a product of three M{\o}ller operators. That it works up to three M{\o}ller operators and not only up to two, is due to the fact that the \(^2P_{3/2}\) interaction is taking place between both \(nc\) and \(n'c\) pairs.
Any inclusion of a fourth M{\o}ller operator based on these two-body interactions would necessarily yield terms in the matrix element that involve the operator product \(\bar{\Omega}_{ij}^\dagger \bar{\Omega}_{ij}^\dagger\), i.e., the same interaction would be applied directly twice without a different interaction in between.
Such contributions to the final-state wave function are unphysical: they are not contained in the multiple-scattering series.

\begin{figure}[H]
    \centering
    \includegraphics[width=0.45\textwidth]{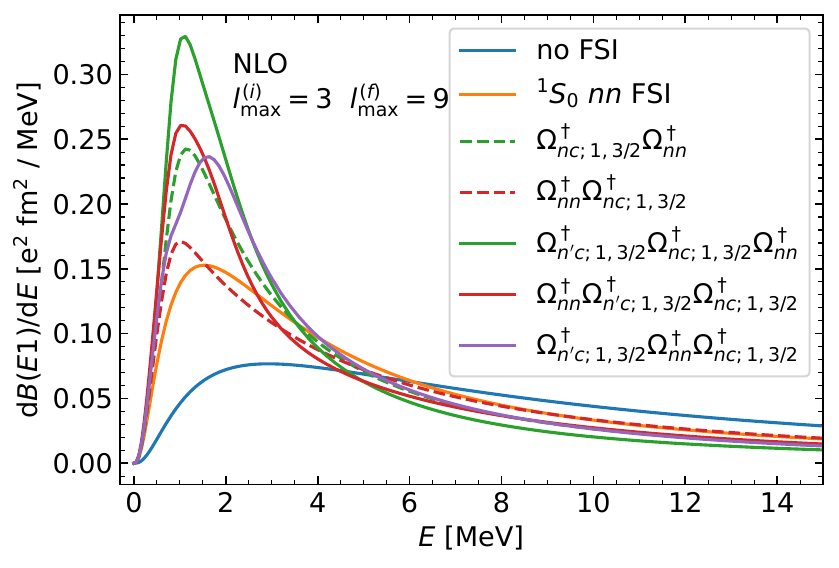}
    \caption{NLO FREFT \(E1\) distributions with different FSI treatments in comparison.
    The green dashed and red dashed curves stem from the application of the product of the \(^1S_0\) \(nn\) and the \(^2P_{3/2}\) \(nc\) M{\o}ller operator. They differ in the order of application.
    The green, purple, and red curves stem from the product of three M{\o}ller operators.
   Green means that \(nn\) FSI is applied first (reading from the right, from the bound state in \cref{eq:e1_distrib_final}) and red means that \(nn\) FSI is applied last (immediately before projection onto the free partial-wave state).}
    For comparison also the distributions without FSI (blue curve) and with \(nn\) FSI only (orange curve) are shown.
    \label{fig:nlo_fsi_so_to}
\end{figure}

At second order, i.e., with products of two M{\o}ller operators, we try the combinations \(\Omega^\dagger_{nc;1,3/2} \Omega^\dagger_{nn}\) and \(\Omega^\dagger_{nn} \Omega^\dagger_{nc;1,3/2}\).
For third order we employ the combinations \(\Omega^\dagger_{n'c;1,3/2} \Omega^\dagger_{nc;1,3/2} \Omega^\dagger_{nn}\), \(\Omega^\dagger_{n'c;1,3/2} \Omega^\dagger_{nn} \Omega^\dagger_{nc;1,3/2}\), and \(\Omega^\dagger_{nn} \Omega^\dagger_{n'c;1,3/2} \Omega^\dagger_{nc;1,3/2}\).
In the case of the \(nc\) M{\o}ller operators we annotated the \(l\) and \(j\) of the partial wave in the subscript for clarity.
Note the difference between \(\Omega^\dagger_{nc;1,3/2}\) and \(\Omega^\dagger_{n'c;1,3/2}\);
these operators differ in the neutron interacting with the core, with the relation 
between them being  \(\Omega^\dagger_{n'c;1,3/2} = \K{-\pmo} \Omega^\dagger_{nc;1,3/2} \K{-\pmo}\). 
The results are shown in \cref{fig:nlo_fsi_so_to}.

We observe that the two combinations of two M{\o}ller operators result in a rather different peak height.
The combination with the \(nn\) FSI applied first\footnote{
    First refers to the view point from the bound state after application of the \(E1\) operator.
    Thereby \(nn\) FSI applied first means that in the product of Hermitian conjugated M{\o}ller operators the \(nn\) operator appears at the right side.
}
yields a peak which is roughly 59\% larger than the one of the curve with \(nn\) FSI only.
In contrast to that, if \(nn\) FSI is applied last the peak height is still larger than in the \(nn\)-FSI-only calculation, but only by about 12\%.
Interestingly, employing \(\Omega^\dagger_{n'c;1,3/2} \Omega^\dagger_{nc;1,3/2}\) (not included in the figure) yields results quite similar to the combination \(\Omega^\dagger_{nc;1,3/2} \Omega^\dagger_{nn}\).
The difference in the peak height is about 1\%, while the peak positions deviate a bit more.
The combination involving \(nn\) FSI peaks around 1.1~MeV in contrast to the pure \(nc\) FSI combination peaking at about 1.5~MeV.

In the case of products of three M{\o}ller operators we observe that for all three curves the peaks are higher than the one of \(nn\) FSI only.
The pattern is similar to that seen for the product of two M{\o}ller operators.
For the combination \(\Omega^\dagger_{n'c;1,3/2} \Omega^\dagger_{nc;1,3/2} \Omega^\dagger_{nn}\) the peak height is about 0.33~\(e^2 \textrm{fm}^2 / \textrm{MeV}\).
This is an increase by approximately 120\% compared to \(nn\) FSI (at NLO) and by about 330\% over the no-FSI distribution.
The other third-order results stemming from the two other positions of the \(nn\) M{\o}ller operator within this product of three operators yield smaller peak heights.
The lowest is obtained by employing \(\Omega^\dagger_{n'c;1,3/2} \Omega^\dagger_{nn} \Omega^\dagger_{nc;1,3/2}\).
It amounts to about 0.24~\(e^2 \textrm{fm}^2 / \textrm{MeV}\).
We checked the fulfillment of the non-energy-weighted sum rule for all three combinations of M{\o}ller operators.
Within the numerical uncertainties the cumulative strength goes as \(E \to \infty\) to the value expected based on \(\expval{r_c^2}\).

We tested also FSI based on truncating the multiple-scattering series at second order (for an explanation see \cref{ssec:fsi}).
The curve is quite similar to the one obtained with \(\Omega^\dagger_{n'c;1,3/2} \Omega^\dagger_{nc;1,3/2} \Omega^\dagger_{nn}\) and leads only to a small violation of the sum rule, although truncating the multiple-scattering series at a particular order is not unitary by construction.
Compared to the product of three M{\o}ller operators the peak is about 3.5\% lower.
The distribution is depicted in Appendix~\ref{ap:mss}.

Thus, while the convergence pattern for the MOPA FSI results could be better, the result based on \(\Omega^\dagger_{n'c;1,3/2} \Omega^\dagger_{nc;1,3/2} \Omega^\dagger_{nn}\) is very close to the one stemming from truncating the multiple-scattering series at second order.
As the comparison with experimental data in Sec.~\ref{ssec:cmp_expm} will show,  it is also closest to the experimental data.
We choose it as our most realistic result.

In the future one could improve on this by calculating the three-body scattering final state for each \(\ket{p,q,\Omega_f}\) boundary condition and employing this for obtaining the \(E1\) strength distribution.

All calculations in this subsection have been carried out with a truncation of the included partial waves given by \(\lmi=3\) and \(\lmf=9\).
For the curve with \(nn\) FSI the estimated uncertainty due to that in the interval up to \(E=15\)~MeV is below 2.5\%. In the case of the combination of three FSIs the uncertainty is estimated to be below 10\% for the same interval.
For more details on the estimates see Appendix~\ref{ap:pw_trunc}.

\subsection{\texorpdfstring{2D \(E1\) strength distribution}{2D E1 strength distribution}}

\begin{figure}[H]
    \centering
    \includegraphics[width=0.8\textwidth]{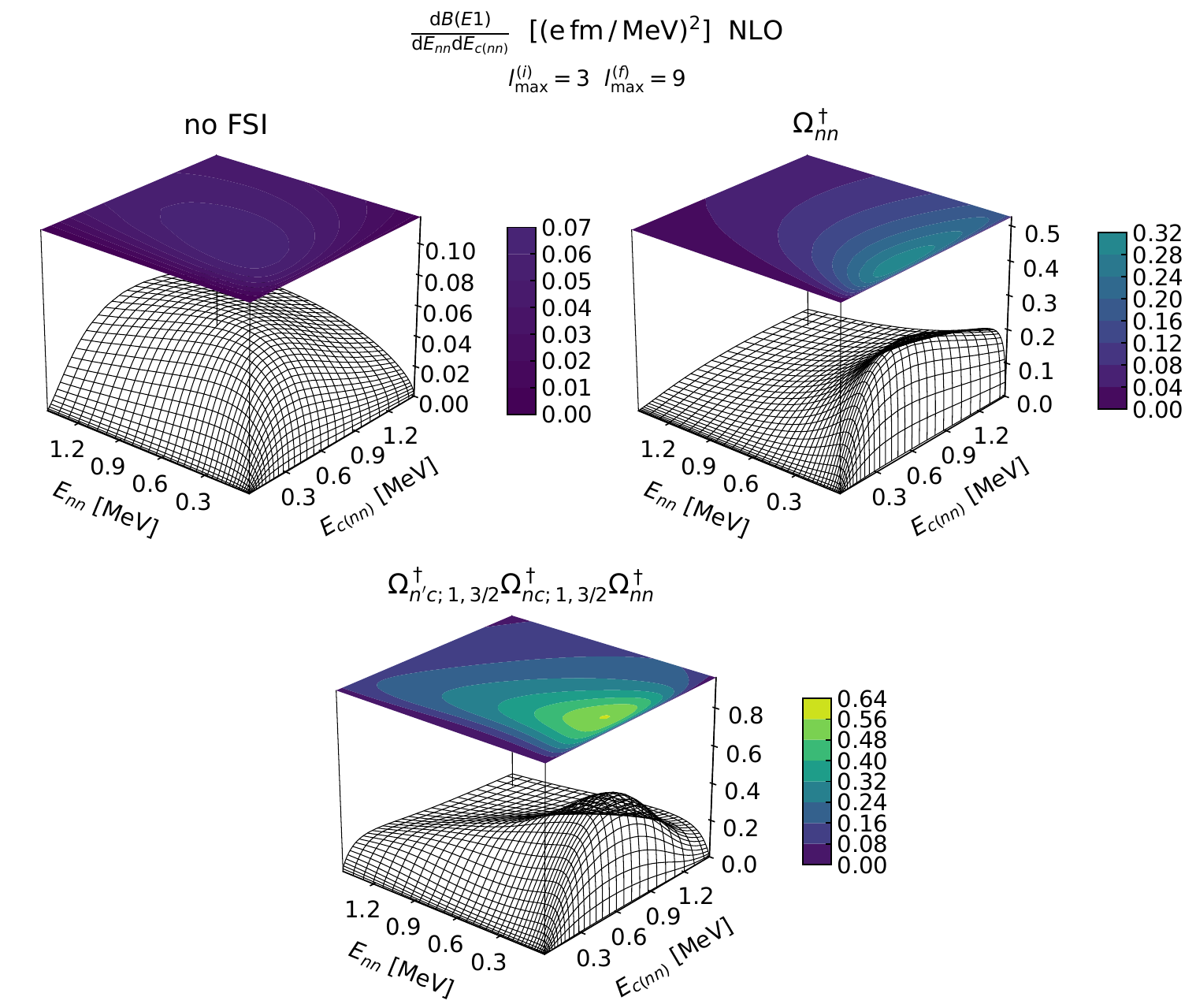}
    \caption{The differential \(E1\) strength \(\dd B(E1) / \K{\dd{E_{nn}} \dd{E_{c(nn)}}}\) at NLO, as a function of the subsystem energies \(E_{nn}\) and \(E_{c(nn)}\).
    The upper left panel shows the result without FSI, in the upper right panel \(nn\) FSI is included.
    In the lower panel the MOPA approach is used, first \(nn\) FSI is applied, then \(nc\) FSI, and finally \(nc\) FSI in the \(n'c\) subsystem.
    Note that the color map is the same for all three panels, only the discretization in terms of levels differs across the panels as is indicated by the individual color bars.
    Note that for compactness the label for z-axes including the unit is part of the annotations in the central area of the plot.}
    \label{fig:2d_e1_distrib}
\end{figure}

Te differential \(E1\) strength, defined as a function of the \(nn\) relative energy, \(E_{nn}\), and the energy of the core relative to the \(nn\) pair, \(E_{c(nn)}\):
\begin{align}\label{eq:2d_e1_distrib_final}
    \frac{\dd B(E1)}{\dd{E_{nn}} \dd{E_{c(nn)}} } = \int {\rm d}p \, {\rm d}q \; q^2 p^2 \frac{1}{2J_i+1} \sum_{\mu,\Omega_f} \left| \sum_{\Omega_i} \imel{c}{p,q;\Omega_f}{\tilde{\Omega} \mathcal{M}_{E1,\mu} P_{\Omega_i}}{\Psi}{} \right|^2  \delta{\K{E_{nn} - \frac{p^2}{2\mu_{nn}} }} \delta{\K{ E_{c(nn)} - \frac{q^2}{2\mu_{c(nn)}} }} \,,
\end{align}
contains more fine-grained information on the role of FSI in the \(E1\) strength.
The two-dimensional distributions without FSI, with \(nn\) FSI, and based on a combination of three M{\o}ller operators are shown in \cref{fig:2d_e1_distrib}.

One can nicely see how the peak position and height changes when one adds FSIs.
The distribution without FSI peaks at around \((490, 510)\), whereby the first number is \(E_{nn}\) in keV and the second number is \(E_{c(nn)}\) in keV.
The peak value is about \(6.9 \times 10^{-2}\) \((\mathrm{e fm}/\mathrm{MeV})^2\).
By adding \(nn\) FSI in the middle panel, the peak is shifted to lower \(E_{nn}\), while the peak position along the \(E_{c(nn)}\) does not seem to change much.
It is now located at \((70, 570)\) and amounts to 0.31~\((\mathrm{e fm}/\mathrm{MeV})^2\).
The peak position along the \(E_{nn}\) axis is not too far from the \(E_{nn}=100\)~keV one would estimate based on the \(nn\) scattering length.
Applying additionally \(nc\) FSI twice (in different \(nc\) pairs) changes the peak height to 0.56~\((\mathrm{e fm}/\mathrm{MeV})^2\) and the position to \((140, 630)\).
We observe also from the contours that the peak structure got more complicated.
Finally, we remark that also in the case of the double-differential distribution we don't see
full convergence in the MOPA FSI scheme. However, our most realistic result is the one given in the right panel
making use of the product of three M{\o}ller operators.

\subsection{Checking the power counting }
\label{ssec:power_counting}

In the previous two subsections we have  discussed the different NLO corrections to the ground-state momentum distribution as well as the inclusion of final-state interactions.
Now we want to check how well the difference between
the LO and NLO results---both on the ground-state level and after the inclusion of FSI---agrees with the expectations from the power counting.
In other words, we want to check whether the Halo EFT power counting is consistent for this observable. 

\begin{figure}[H]
    \centering
    \includegraphics[width=0.45\textwidth]{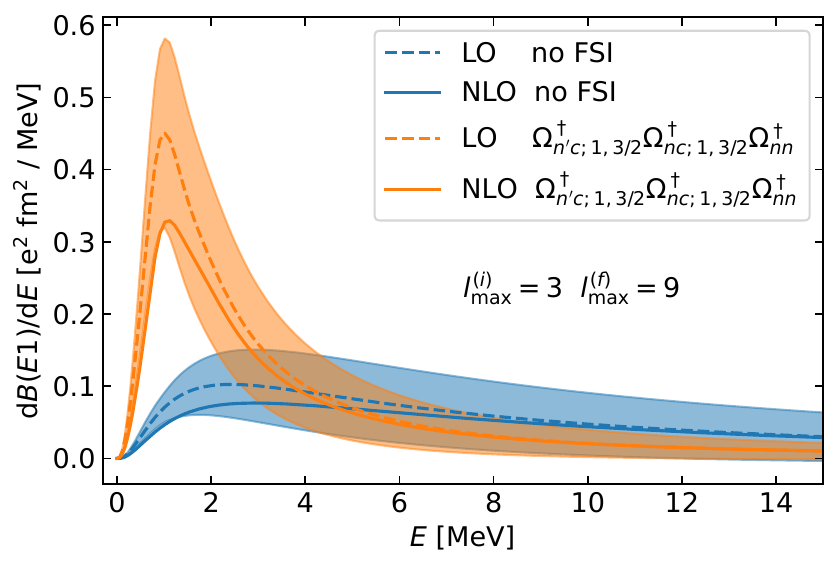}
    \caption{LO (dashed) and NLO (solid) FREFT \(E1\) distributions with (orange) and without (blue) FSIs included.
    The results with FSI included are based on the product of M{\o}ller operators \(\Omega_{n'c;1,3/2}^\dagger \Omega_{nc;1,3/2}^\dagger \Omega_{nn}^\dagger\), thereby including both \(^1S_0\) \(nn\) FSI and \(^2P_{3/2}\) \(nc\) FSI.
    The uncertainty bands indicate the LO EFT uncertainties calculated using \cref{eq:unc_frac}.}
    \label{fig:ver_pc}
\end{figure}

The expectation from the power counting is that NLO corrections are approximately given by the following fraction being a function of the energy:
\begin{equation}\label{eq:unc_frac}
    u{(E)} = \max{\K{\K{S_{2n}/E_{\mathrm{hi}}}^n, \K{E/E_{\mathrm{hi}}}^n}} \,,
\end{equation}
whereby \(E_{\mathrm{hi}}\) is the high-energy scale given converting the high-momentum scale in the \(nc\) system to an energy scale using the approximate relation  \(E_{\mathrm{hi}} \approx (150~\mathrm{MeV})^2 / (2 m_n) \approx 12\)~MeV.
For final-state energies below the two-neutron separation energy, the uncertainty as a fraction is given by this energy over the high-energy scale.
At higher final energies the uncertainty is determined by the energy over the high-energy scale.
The exponent \(n\) is one half for LO and one for NLO.
\Cref{fig:ver_pc} compares the LO results (shown as dashed curves) with the NLO results (shown as solid curves). It displays a fractional uncertainty band according to the formula to aid the comparison.
The ground-state curves are shown in blue together with orange curves including both \(^1S_0\) \(nn\) FSI and \(^2P_{3/2}\) \(nc\) FSI by using \(\Omega_{n'c;1,3/2}^\dagger \Omega_{nc;1,3/2}^\dagger \Omega_{nn}^\dagger\).
Note that the  FSI calculations for the LO result use the LO FSI M{\o}ller operators and those for the NLO result use the NLO FSI M{\o}ller operators. 
As we already discussed, in our FREFT calculation the LO and NLO \(^2P_{3/2}\) \(nc\) are the same, so the difference between the FSIs stems from  the finite effective range in the NLO \({}^1S_0\) \(nn\) amplitude. 
In principle at NLO there is also the FSI \(^2S_{1/2}\) \(nc\) interaction, that was discussed in the previous subsection. Here it is neglected for simplicity, since it is comparatively weak.

We observe that the NLO result without FSI is substantially lower than the LO ground-state momentum distribution, but it is mainly within the uncertainty band of the LO result.
Only around energies of 1~MeV it is slightly outside this band.
The NLO result with FSI is also substantially lower than the corresponding LO calculation.
Also in this case the NLO result is mainly within the uncertainty band of the LO result, with again a slight exemption around 1~MeV.
We conclude that the power counting works as expected.

\subsection{Final result and comparison with experimental data}
\label{ssec:cmp_expm}

Our final result is then the solid orange curve of Fig.~\ref{fig:ver_pc} (being the same as the solid green curve in Fig.~\ref{fig:nlo_fsi_so_to}). It uses the initial-state wave function with the NLO effects incorporated as described above. The ${}^2P_{3/2}$ $n \alpha$ and ${}^1S_0$ $nn$ FSIs are both included using FREFT M{\o}ller operators. Specifically, the combination \(\Omega_{n'c;1,3/2}^\dagger \Omega_{nc;1,3/2}^\dagger \Omega_{nn}^\dagger\) is used.

However, this curve has a substantially smaller uncertainty band than that associated with the LO results in the previous subsection. The uncertainty due to NNLO mechanisms in the ${}^6$He wave function is estimated using \cref{eq:unc_frac} with \(n=1\). It is indicated by the orange band. Note that band only accounts for the uncertainty due to the interactions of the initial-state calculation and of the final-state calculation. There are additional uncertainties due to the approximate FSI scheme and the truncation of the sum over partial waves, these are not quantified here.

\begin{figure}[H]
    \centering
    \includegraphics[width=0.45\textwidth]{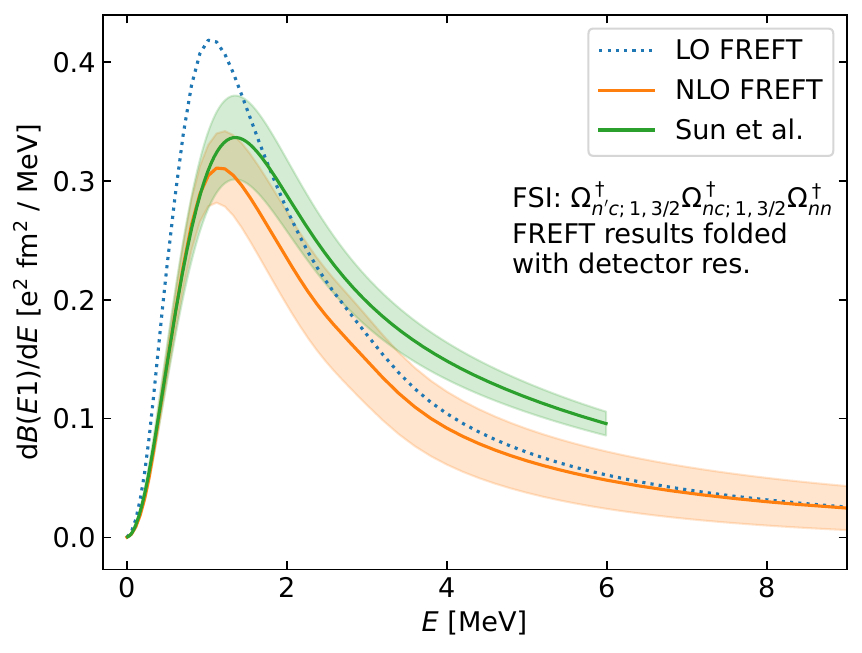}
    \caption{\label{fig:e1_fr}The NLO FREFT \(E1\) distribution folded with the detector resolution (in orange) in comparison with the band for the E1 strength distribution extracted from experimental data by Sun et al. \cite{Sun:2021efo} (in green). 
    The uncertainty band around the NLO result indicates the EFT uncertainty calculated using \cref{eq:unc_frac} (\(n=1\)). Uncertainties of the
    FSI treatment are not included. For comparison, we also show the LO result in blue. Its uncertainty is not shown here.}
\end{figure}

Fig.~\ref{fig:e1_fr} also shows the LO result (with corresponding LO FSI calculations) in blue. Both the LO and NLO results shown in \cref{fig:e1_fr} are folded with the detector resolution quoted in Ref.~\cite{Sun:2021efo}. We  also include the band for the $E1$ strength distribution extracted from data on the Coulomb dissociation of ${}^6$He by Sun \textit{et al.}~\cite{Sun:2021efo}. 

We observe that NLO result with its uncertainty band and the experimental data agree within their respective uncertainty bands up to energies of roughly 3~MeV, which is already beyond the peak position.
Around \(E=5\)~MeV there are considerable deviations.
The leading-order result in general doesn't perform better in comparison with the experimental data.
Its consistency with the NLO result within the uncertainties expected based on the power counting
was already seen in the previous subsection.


\section{Summary and Outlook}
\label{sec:conclusion}

We have implemented a three-body calculation in Halo Effective Field Theory (Halo EFT) with finite-range potentials. The finite-range EFT (FREFT) developed in this paper uses momentum-space one-term separable interactions with Yamaguchi form factors. We followed Ref.~\cite{Li:2023hwe} and argued that these potentials provide a way to realize the Halo EFT hierarchy of mechanisms in the $nn$ and $n \alpha$ amplitudes. Li \textit{et al.}~\cite{Li:2023hwe} and Contessi \textit{et al.}~\cite{Contessi:2023yoz} point out that when finite-range interactions reproduce the necessary terms in the effective-range expansion in both these waves, and a three-body force is included in the calculation, FREFT  produces results equivalent to an infinite-cutoff formulation of the EFT at the leading-order accuracy of such a calculation.  FREFT uses finite-range interactions  that are energy independent  and correspond to Hermitian Hamiltonians. It therefore provides a practical way to calculate two-body, three-body, \ldots observables in a quantum-mechanical formulation. 

We computed the ${}^6$He ground state at leading order and next-to-leading order in this approach, and obtained the $E1$ strength of the nucleus up to energies of order 10 MeV. This is, to our knowledge, the first calculation of ${}^6$He to NLO in Halo EFT (see Ref.~\cite{Thapaliya:2016knc} for an earlier attempt in a formulation of the problem with energy-dependent interactions). Note that in contrast to Contessi \textit{et al.}'s finite-range implementation of EFT in few-body systems we did not evaluate NLO corrections in strict perturbation theory, instead modifying the kernel of our three-body integral equation to account for the additional effects in $nn$ and $n \alpha$ scattering that occur at NLO. 

Using the ground-state description we calculated the distance of the \(\alpha\) core from the center of mass. Based on that and on literature data we obtained the charge radius of \(^6\)He, finding
2.06~fm at LO and 2.00~fm at NLO. We estimate the EFT uncertainty to be 0.35 fm at LO and 0.09 fm at NLO. 
The Halo EFT result for the  \(^6\)He charge radius is thus in good agreement with the experimental value of \(2.0571 \pm 0.0075\)~fm \cite{Krauth:2021foz}.

Final-state interactions were included in the \(E1\) strength calculation using an approximate scheme based on M{\o}ller operators. We found that the neutron-neutron FSI is the largest effect at low energies, with the \(^2P_{3/2}\) core-neutron FSI having a slightly smaller but still large effect.
This is in contrast to the observations for \(^{11}\)Li where the effect of the \(s\)-wave \(nc\) interaction was much weaker than the one of the \(nn\) interaction \cite{Gobel:2022pvz}.
Both interactions can interfere constructively, so our final results accounted for both, through the multiplication of a separate M{\o}ller operator for each of these FSIs.
We used a product of three M{\o}ller operators to account for the \(nc\) interaction in the \(nc\) as well as in \(n'c\) (neutrons interchanged) subsystem.
As an alternative, we tested truncating the multiple-scattering series in the t-matrices at second order which yielded a quite similar result.

The NLO correction to the ${}^6$He $E1$ strength is relatively large in the peak of the $E1$ energy distribution (\(E=1-2\) MeV) where it is roughly 30\%. It becomes smaller at energies of 7-10 MeV. The size of the NLO correction is therefore consistent with the expected expansion parameter of Halo EFT for ${}^6$He.

We compared our $E1$ strength distribution computation with the experimental data of Sun \textit{et al.}~\cite{Sun:2021efo}. While the NLO calculation produces the data reasonably well up to final-state energies of about 3~MeV, at higher energies the deviations exceed the sum of the experimental and theory uncertainties.
Future work might help understand this discrepancy better.
In particular, it would be interesting to base future FSI calculations on a full three-body calculation of the final scattering states by means of the Faddeev equations for continuum states.

We also have access to differential $E1$ distributions in our approach, which allows comparison with more fine-grained experimental information on the distribution of the three particles in the final state. And the NLO finite-range EFT calculation carried out here can now be employed to check the model-based argument of Ref.~\cite{Gobel:2021pvw} that NLO corrections to the structure of ${}^6$He do not vitiate the possibility to extract the neutron-neutron scattering length from the process ${}^6$He$(p,p'\alpha)nn$. 

More generally, finite-range EFT facilitates computations of the structure of halo nuclei in Halo EFT beyond leading order. In the case of ${}^6$He it raises the possibility that a  calculation at next-to-next-to-leading order (N2LO), where the amplitude for the ${}^2P_{1/2}$ channel enters, could be carried out. Corrections to the ${}^2P_{3/2}$ channel associated with the \(p\)-wave ``shape parameter'' would also be needed at N2LO. The FREFT potentials we used here would therefore have to be generalized to produce this additional term in the effective-range expansion. Future research on FREFT will be needed to determine how general the finite-range implementation must be in order to permit all values of the \(p\)-wave shape parameter that are consistent with the generalization of the Wigner bound~\cite{Wigner:1955zz} to \(p\)-waves~\cite{Hammer:2009zh}. Finally, whether an additional three-body force is needed at N2LO in ${}^6$He, as it is in the case of halo nuclei with purely \(s\)-wave interactions~\cite{Bedaque:2002yg}, will be an important aspect of such an N2LO calculation.

\acknowledgments

We thank T. Aumann, T. Nakamura, and Y. Sun for useful discussions.
Moreover, we thank T. Nakamura and Y. Sun for sharing the experimental data with us.
H.W.H. acknowledges support by Deutsche Forschungsgemeinschaft (DFG, German Research Foundation) - Project-ID 279384907 - SFB 1245 and the BMFTR under grant no. 05P24RDB.
D.R.P. was supported by the US Department of Energy, contract DE-FG02-93ER40756 and by the Swedish Research Council via a Tage Erlander Professorship (Grant No 2022-00215).
M.G. acknowledges support by the Czech Science Foundation GA\scv{C}R Grant No. 25-18335S.


\newpage 

\appendix

\section{\texorpdfstring{Antisymmetrization operator \(A_{ij}\) for the Faddeev equations}{Antisymmetrization operator Aij for the Faddeev equations}}
\label{ap:Aij}

The definition of the operator for writing the antisymmetrized Faddeev equations reads
\begin{align}
    A_{ij} \coloneqq 
        \begin{cases}
            0         & \textrm{for } i   =  j \land \mathcal{S}{(j)} = c \,, \\
            -\pmo     & \textrm{for } i   =  j \land \mathcal{S}{(j)} = n \,, \\
            \id       & \textrm{for } i \neq j \land \mathcal{S}{(j)} = c \,, \\
            \id -\pmo & \textrm{for } i \neq j \land \mathcal{S}{(j)} = n \,. \\
        \end{cases}
\end{align}

\section{\texorpdfstring{Matrix elements of \(E1\) operator in arbitrary partial waves}{Matrix elements of E1 operator in arbitrary partial waves}}
\label{ap:E1_mel}

We present matrix elements of the \(E1\) operator between arbitrary partial-wave states in coordinate- and momentum space for different coupling schemes.
They can be seen as intermediate steps in a derivation of the momentum-space matrix element in the \(jI\) basis.
We first give an overview of this procedure and afterwards state the results.
One can first evaluate the matrix element in coordinate space in a partial-wave basis with definite \(L\) and \(S\) (not coupled to definite \(J\)).
Due to the nature of the \(E1\) operator with its first spherical harmonic, one then needs also an identity for the complete angular integral over three spherical harmonics, which can be found, e.g., in Ref. \cite{Khersonskii:1988krb} (eq. (4) of chapter 5.9.1).
Based on this, one can derive the expression for the coordinate-space matrix element in a coupled \(LS\) basis.
To write it more compactly sums over products of Clebsch-Gordan coefficients can be expressed in terms of Wigner-6j symbols.
The recoupling from a coupled \(jI\) basis to a coupled \(LS\) basis is done via Wigner-9j symbols.
Finally, one can obtain the matrix element in a \(jI\) basis.
A useful expression for the recoupling can be found in Ref. \cite{Ji:2014wta}.
To simplify a resulting sum of products of Wigner-9j and Wigner-6j symbols one can consult Ref. \cite{Khersonskii:1988krb} (eq. (41) of chapter 12.2.4).
The result for the matrix element in our \(jI\) coupling scheme has then to be transformed from coordinate space to momentum space, whereby
\begin{equation}
    \rint{\x} j_l{\K{p'x}} j_l{\K{px}} = \frac{\pi}{2} \frac{\delta{\K{p'-p}}}{p'^2}
\end{equation}
as well as identities relating spherical Bessel functions of different order via derivatives are employed.

We start with the coordinate-space matrix elements.
In a plane-wave basis the expression reads
\begin{align}\label{eq:mel_e1_op_vec_basis}
    \imel{c}{\vxyp}{r_c \y{1}{\mu}{\v{r_c}}}{\vxy}{c} &= f_c y \y{1}{\mu}{\vy} \ibraket{c}{\vxyp}{\vxy}{c} \\
    &= f_c y \y{1}{\mu}{\vy} \dt{\vxp-\vx} \dt{\vyp-\vy} \,.
\end{align}

In an uncoupled \(LS\) basis we have
\begin{align}
    & \imel{c}{\xyp;(l',\lambda')L',M'_L;(s',\sigma')S',M'_S}{ r_c \y{1}{\mu}{r_c} }{\xy;(l,\lambda)L,M_L;(s,\sigma)S,M_S}{c} \nonumber \\
    &\quad = \kd{s}{s'} \kd{\sigma}{\sigma'} \kd{S}{S'} \kd{M_S}{M'_S} \kd{l}{l'} \frac{\delta{\K{x'-x}}}{x'^2} \frac{\delta{\K{y'-y}}}{y'^2} \sqrt{\frac{3}{4\pi}} f_c y \sqrt{\hat{L} \hat{\lambda}}  \nonumber \\
    &\quad \quad \times \cgcc{\lambda,0,1,0}{\lambda',0} \cgcc{L,M_L,1,\mu}{L',M'_L} \K{-1}^{\lambda+l+L'+1} \begin{Bmatrix} \lambda & l & L \\ L' & 1 & \lambda' \end{Bmatrix} \,.
\end{align}

For the coupled \(LS\) basis the matrix element is given by
\begin{align}
    & \imel{c}{\xyp;(l',\lambda')L'(s',\sigma')S';J',M'}{  r_c \y{1}{\mu}{r_c}  }{\xy;(l,\lambda)L(s,\sigma)S;J,M}{c} 
      \nonumber \\
    &\quad = \kd{s}{s'} \kd{\sigma}{\sigma'} \kd{S}{S'} \kd{l}{l'} \frac{\delta{\K{x'-x}}}{x'^2} \frac{\delta{\K{y'-y}}}{y'^2} \sqrt{\frac{3}{4\pi}} f_c y \sqrt{\hat{L}' \hat{L}} \sqrt{\hat{J} \hat{\lambda}} \nonumber \\
    &\quad \quad \times \K{-1}^{\lambda+l+S+J} \cgcc{\lambda,0,1,0}{\lambda',0} \cgcc{J,M,1,\mu}{J',M'} \begin{Bmatrix} L & S & J \\ J' & 1 & L' \end{Bmatrix} \begin{Bmatrix} \lambda & l & L \\ L' & 1 & \lambda' \end{Bmatrix} \,.
\end{align}

In the \(jI\) basis one has
\begin{align}
    & \imel{c}{\xyp;\Omega'}{r_c \y{1}{\mu}{r_c}}{\xy;\Omega}{c} \nonumber \\
    &\quad = \kd{s}{s'} \kd{\sigma}{\sigma'} \kd{l}{l'} \kd{j}{j'} \frac{\delta{\K{x'-x}}}{x'^2} \frac{\delta{\K{y'-y}}}{y'^2} \sqrt{\frac{3}{4\pi}} f_c y \sqrt{\hat{I} \hat{I'}} \sqrt{\hat{\lambda} \hat{J}} \nonumber \\
    &\qquad \times \cgcc{\lambda,0,1,0}{\lambda',0} \cgcc{J,M,1,\mu}{J',M'} \K{-1}^{2s + \sigma + \lambda' + j'} \K{-1}^{-J'} 
      \begin{Bmatrix} 1 & I' & I \\ j' & J & J'\end{Bmatrix} \begin{Bmatrix} 1 & I' & I \\ \sigma & \lambda & \lambda' \end{Bmatrix} 
      \,.
\end{align}

The momentum-space relation in the uncoupled \(LS\) basis is
\begin{align}
    & \imel{c}{\pqp;(l',\lambda'=\lambda \pm 1)L',M'_L;(s',\sigma')S',M'_S}{ r_c \y{1}{\mu}{r_c} }{\pq;(l,\lambda)L,M_L;(s,\sigma)S,M_S}{c} \nonumber \\
    &\quad = \kd{s}{s'} \kd{\sigma}{\sigma'} \kd{S}{S'} \kd{M_S}{M'_S} \kd{l}{l'} \ci^{\lambda'-\lambda} \frac{\delta{\K{p'-p}}}{p'^2} D_{q'}^{\K{\lambda, \pm 1} } \frac{\delta{\K{q'-q}}}{q'^2} \sqrt{\frac{3}{4\pi}} f_c \sqrt{\hat{L} \hat{\lambda}}  \nonumber \\
    &\quad \quad \times \cgcc{\lambda,0,1,0}{\lambda',0} \cgcc{L,M_L,1,\mu}{L',M'_L} \K{-1}^{\lambda+l+L'+1} \begin{Bmatrix} \lambda & l & L \\ L' & 1 & \lambda' \end{Bmatrix} \,,
\end{align}
whereby the definition of \(D_{q'}^{\K{\lambda, d} }\) can be found in \cref{eq:def_dop}.

In the coupled \(LS\) basis it is given by
\begin{align}
    & \imel{c}{\pqp;(l',\lambda'=\lambda \pm 1)L'(s',\sigma')S';J',M'}{  r_c \y{1}{\mu}{r_c}  }{\pq;(l,\lambda)L(s,\sigma)S;J,M}{c} 
      \nonumber \\
    &\quad = \kd{s}{s'} \kd{\sigma}{\sigma'} \kd{S}{S'} \kd{l}{l'} \ci^{\lambda'-\lambda}  \frac{\delta{\K{p'-p}}}{p'^2} D_{q'}^{\K{\lambda, \pm 1} } \frac{\delta{\K{q'-q}}}{q'^2} \sqrt{\frac{3}{4\pi}} f_c \sqrt{\hat{L}' \hat{L}} \sqrt{\hat{J} \hat{\lambda}} \nonumber \\
    &\quad \quad \times \K{-1}^{\lambda+l+S+J} \cgcc{\lambda,0,1,0}{\lambda',0} \cgcc{J,M,1,\mu}{J',M'} \begin{Bmatrix} L & S & J \\ J' & 1 & L' \end{Bmatrix} \begin{Bmatrix} \lambda & l & L \\ L' & 1 & \lambda' \end{Bmatrix} \,.
\end{align}

The final momentum-space expression for the \(jI\) basis is already given in \cref{eq:ms_jI_mel_E1_op}.

\section{Expression for the recoupling operation}
\label{ap:recpl}

The analytically simplified expression for the transition to a different spectator reads
\begin{align}
    &\rint{\pp} \rint{\qp} \ibraket{\mathcal{S}{(j)}}{p,q;\Omega}{\pp, \qp; \Omega'}{\mathcal{S}{(i)}} f{\K{\pp, \qp}} \nonumber \\
  &= \kd{J}{J'} \kd{M}{M'} \sum_{L,S} \sqrt{\hat{j} \hat{I} \hat{j}' \hat{I}'} \hat{S} \hat{L}
    \begin{Bmatrix} l & s & j \\ \lambda & \sigma & I \\ L & S & J \end{Bmatrix}
    \begin{Bmatrix} l' & s' & j' \\ \lambda' & \sigma' & I' \\ L & S & J \end{Bmatrix} 
    \ibraket{\mathcal{S}{(j)}}{(s,\sigma)S,M_S=S}{(s',\sigma')S,M_S=S}{\mathcal{S}{(i)}} \nonumber \\
  &\quad \times \sum_{l'_1+l'_2 = l'} \tilde{F}_{jip}^{\K{l'_1,l'_2;l'}}{\K{p,q}}
  \sum_{\lambda'_1+\lambda'_2 = \lambda'} \tilde{F}_{jiq}^{\K{\lambda'_1,\lambda'_2;\lambda'}}{\K{p,q}}  \sum_{\lt} \frac{1}{2} g_{ji;\lt}^{(l',\lambda')}{\K{p,q}} 
  \sum_{\Lt,\Lt'} \hat{\lt} \sqrt{\hat{l}\hat{\lambda}\hat{l}'_1 \hat{l}'_2 \hat{\lambda}'_1 \hat{\lambda}'_2 \hat{l}' \hat{\lambda}'} \nonumber \\
  &\quad \times \cgcc{l,0,\lt,0}{\Lt,0} \cgcc{\lambda,0,\lt,0}{\Lt',0} \cgcc{l'_1,0,\lambda'_1,0}{\Lt,0} \cgcc{l'_2,0,\lambda'_2,0}{\Lt',0} \K{-1}^{l+L+\tilde{L}'}
   \begin{Bmatrix} L & \Lt & \Lt' \\ \lt & \lambda & l \end{Bmatrix}
   \begin{Bmatrix} L & l' & \lambda' \\ \Lt & l'_1 & \lambda'_1 \\ \Lt' & l'_2 & \lambda'_2 \end{Bmatrix} \,,
   \label{eq:ovl_fnl}
\end{align}
whereby 
\begin{align}
  g_{ji}^{(l',\lambda')}{\K{p,q,x}} &\coloneqq \left| \frac{\kjip{p,q,x=0}}{\kjip{\vpq}} \right|^{l'} \left| \frac{\kjiq{p,q,x=0}}{\kjiq{\vpq}} \right|^{\lambda'}
  f{\K{\kjip{\K{\vpq}}, \kjiq{\K{\vpq}}}} \,, \\
  g_{ji;\lt}^{(l',\lambda')}{\K{p,q}} &\coloneqq \int \dd{x} P_{\lt}{\K{x}} g_{ji}^{(l',\lambda')}{\K{p,q,x}} \,,
\end{align}
with \(P_l\) being the \(l\)-th Legendre polynomial, and
\begin{align}
  \tilde{F}_{jik}^{\K{l'_1, l'_2; l'}}{\K{p,q}} \coloneqq \frac{ \K{a_{jik} p}^{l'_1} \K{b_{jik} q}^{l'_2} }{ \left| \kjiq{p,q,x=0} \right|^{l'} }
    \sqrt{\frac{\hat{l}'!}{\hat{l}'_1! \hat{l}'_2!}} \, .
\end{align}
The definitions of the \(a_{ijk}\) and \(b_{ijk}\) (with \(k \in \{p, q\}\)) follow from the relations
\begin{align}
    \kjipv{\vpq} &\eqqcolon a_{jip} \v{p} + b_{jip} \v{q} \,, \\
    \kjiqv{\vpq} &\eqqcolon a_{jiq} \v{p} + b_{jiq} \v{q} \,.
\end{align}
The definitions for momentum transformation functions \(\v{\kappa}_{jik}{\K{\vpq}}\) can be found, e.g., in Refs.~\cite{Gobel:2019jba,Gobel:2024ovk}.

Note that \cref{eq:ovl_fnl} corresponds to Eq. (9.101) of Ref.~\cite{Gobel:2024ovk},
whereby the following typos have been corrected.
The meaningless \(\kd{L}{L'} \kd{M_L}{M'_L}\) has been removed and the \(\ibraket{\mathcal{S}{(j)}}{(s,\sigma)S,M_S}{(s',\sigma')S',M'_S}{\mathcal{S}{(i)}}\) has been replaced by \(\ibraket{\mathcal{S}{(j)}}{(s,\sigma)S,M_S}{(s',\sigma')S,M_S}{\mathcal{S}{(i)}}\) (respectively \(\ibraket{\mathcal{S}{(j)}}{(s,\sigma)S,M_S=S}{(s',\sigma')S,M_S=S}{\mathcal{S}{(i)}}\)). These changes are obvious from looking at Eq. (9.100) of that reference. Moreover, \(g_{ji;\lt}^{(l',\lambda)}\) has been replaced by \(g_{ji;\lt}^{(l',\lambda')}\) and the occurrences of functions of the form \(F_{jik}^{\K{l_1,l_2;l}}{\K{p,q}}\) have effectively been corrected into functions of the form \(F_{jik}^{\K{l_1,l_2;l}}{\K{p,q}} / \sqrt{4\pi}\).
Finally, the \(\Lt'\) in the exponent of the \((-1)\) of Eq.~(9.101) has to be replaced by \(\Lt\).
In \cref{eq:ovl_fnl} this long phase factor has been simplified by using the properties of the Wigner-9j symbol (leading to the emergence of among others \(\Lt'\) in the phase factor after the simplifications).

\section{\texorpdfstring{Calculating \(nc\) FSI}{Calculating nc FSI}}
\label{ap:nc_fsi}

The details of how the nontrivial contribution from the \(nc\) FSI is evaluated are already explained in \cref{ssec:fsi}.
The purpose of this appendix is to present some intermediate steps in obtaining \cref{eq:nc_fsi_mel}:
\begin{align}
    &\imel{c}{p,q;\Omega}{ \K{\Omega_{nc} - \id}^\dagger \mathcal{M}_{E1,\mu} }{\Psi}{} \nonumber \\
    &\quad= \ibra{c}{p,q;\Omega} \rint{\pp} \rint{\qp} \sum_{\substack{\Omega' \textrm{ with}\\{\K{\Omega'}_{nc}=\omega_{nc}}}}\iketbra{p',q';\Omega'}{n}{} t_{nc}{\K{E_{p'}}} G_0^{\K{nc}}{\K{E_{p'}}} \mathcal{M}_{E1,\mu} \iket{\Psi}{} \\
    &\quad= \sum_{\Omega_i} \sum_{\substack{\Omega' \textrm{ with}\\{\K{\Omega'}_{nc}=\omega_{nc}}}} \rint{\pp} \rint{\qp} \ibraket{c}{p,q;\Omega}{p',q';\Omega'}{n}  \ibra{n}{p',q';\Omega'} t_{nc}{\K{E_{p'}}} G_0^{\K{nc}}{\K{E_{p'}}} \nonumber \\
    &\quad \quad \times \rint{\ppp} \rint{\qpp} \iket{\ppp,\qpp;\Omega_i}{c} f_{E1,\mu}^{\K{\Omega_m,\Omega_i}} D_{q''}^{\K{\Omega_m,\Omega_i}} \Psi_{c,\Omega_i}{\K{\ppp,\qpp}} \\
    &\quad= \sum_{\Omega_i,\Omega_m} f_{E1,\mu}^{\K{\Omega_m,\Omega_i}} \sum_{\substack{\Omega' \textrm{ with}\\{\K{\Omega'}_{nc}=\omega_{nc}}}} \rint{\pp} \rint{\qp} \ibraket{c}{p,q;\Omega}{p',q';\Omega'}{n}  \nonumber \\
    &\quad \quad \times g_{l_n}{\K{p'}} \tau_{nc}{\K{E_{p'}}} \rint{\pt} g_{l_n}{\K{\pt}} \frac{1}{E_{p'}- \pt[2]/\K{2\mu_{nc}} + \ci \epsilon}  \nonumber \\
    &\quad \quad \times \rint{\ppp} \rint{\qpp} \ibraket{n}{\pt,q';\Omega'}{\ppp,\qpp;\Omega_i}{c} D_{q''}^{\K{\Omega_m,\Omega_i}} \Psi_{c,\Omega_i}{\K{\ppp,\qpp}}
\end{align}
From this expression by making use of the definition given in \cref{eq:O_op} one arrives at \cref{eq:nc_fsi_mel}.

\section{\texorpdfstring{Additional \(E1\) distributions based on single FSIs}{Additional E1 distributions based on single FSIs}}
\label{ap:add_E1_sg_fsis}

In this appendix, we show a little more details regarding specific FSIs and their implementation.
First, we will compare the finite-range and zero-range implementations of the NLO \(^1S_0\) \(nn\) FSI and the \(^2P_{3/2}\) \(nc\) FSI.
Then, we will evaluate different implementations of the \(^2S_{1/2}\) \(nc\) FSI, which enters at NLO.

\begin{figure}[H]
    \centering
    \includegraphics[width=0.92\textwidth]{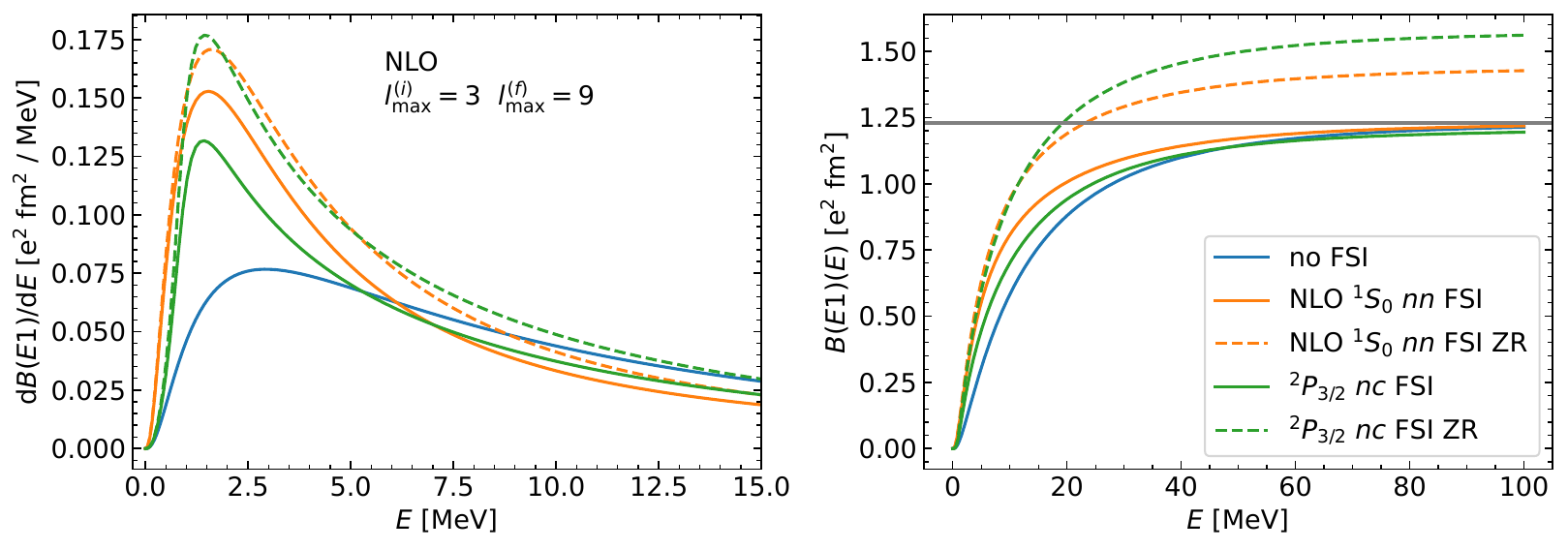}
    \caption{\label{fig:nlo_fsi_cmp1b}\textit{Left panel:} The orange dashed curve is the FREFT \(E1\) distribution but with an NLO $nn$ FSI computed with a zero-range interaction. It is to be contrasted to the orange solid curve, in which a finite-range FSI interaction was employed. The green dashed and green solid curve make a similar comparison for the case of the leading \(nc\) FSI, that in the ${}^2P_{3/2}$ $nc$ channel. 
    \textit{Right panel:} The corresponding cumulative \(E1\) distributions. The gray line shows the expected value for \(B{(E1)}{(E \to \infty)}\) obtained from \(\expval{r_c^2}\) given by the form factor.}
\end{figure}

Figure \ref{fig:nlo_fsi_cmp1b} contrasts calculations that include the
 \(^2P_{3/2}\) \(nc\) FSI and the NLO \(nn\) FSI via M{\o}ller operators computed with zero-range interactions with the corresponding (green and orange) curves from Fig.~\ref{fig:nlo_fsi_cmp1}. The $nn$ and $nc$ amplitudes employed to obtain the dashed curves, i.e., the zero-range results, of \cref{fig:nlo_fsi_cmp1b} correspond to energy-dependent interactions. The use of a zero-range M{\o}ller operator leads to FSI effects that are larger than those in the FREFT calculation. The right panel shows that there is a corresponding overshoot of the sum-rule value of the cumulative \(E1\) strength. This is presumably because the correction terms due to the energy dependence that are relevant for probability conservation have not been taken into account. In contrast to that, the finite-range FSIs fulfill the sum rule.

\begin{figure}[H]
    \centering
    \includegraphics[width=0.92\textwidth]{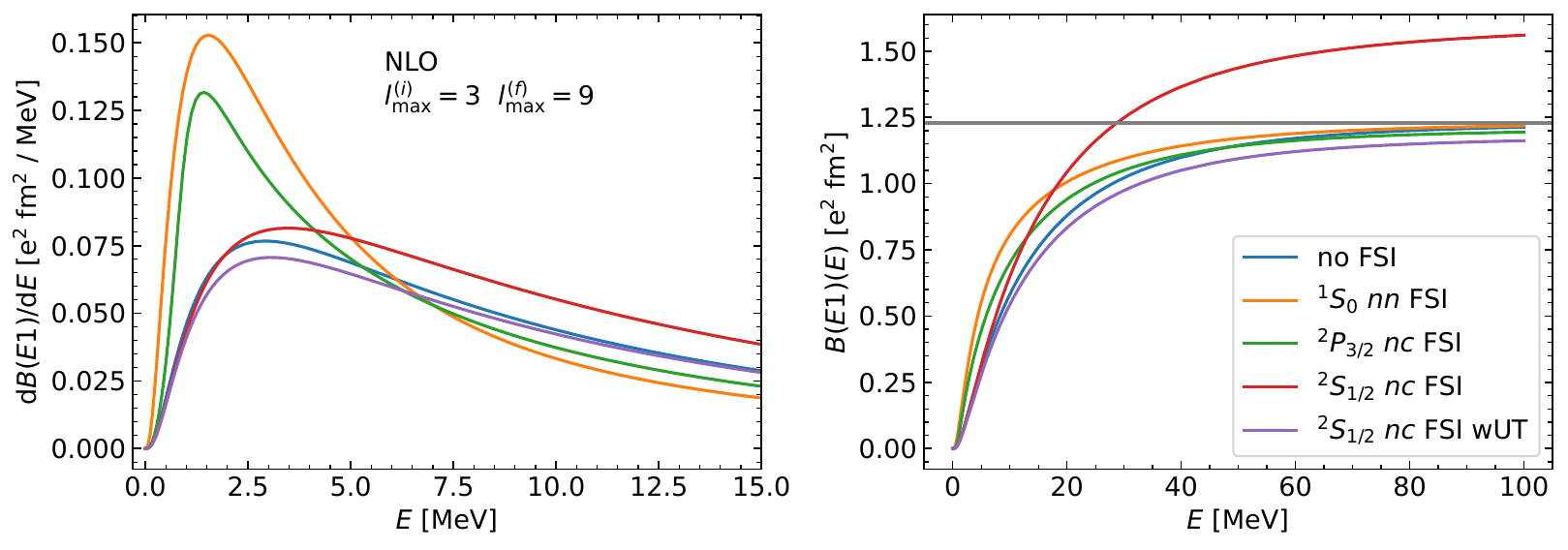}
    \caption{\label{fig:nlo_fsi_cmp2}\textit{Left panel:} The blue curve again represents the FREFT \(E1\) distributions without FSI. The orange and green curves are the results with $nn$ ${}^1S_0$ and $n \alpha$ ${}^2P_{3/2}$ FSI. Also shown in red and purple are two implementations of the \({}^2S_{1/2}\) \(nc\) FSI. The red line is the result with a constant ${}^2S_{1/2}$ $nc$ amplitude, while the purple line shows the modification of this result when the $nc$ unitarity term is included in the amplitude. 
    \textit{Right panel:} The corresponding cumulative \(E1\) distributions. The gray line shows the expected value for \(B{(E1)}{(E \to \infty)}\) obtained from \(\expval{r_c^2}\) given by the form factor. Note that the constant-amplitude implementation of the $n \alpha$ s-wave FSI overshoots the sum rule.
    The implementation with the unitarity term undershoots it slightly.}
\end{figure}

Figure \ref{fig:nlo_fsi_cmp2} shows the effect on the $E1$ strength function of including the ${}^2S_{1/2}$ \(nc\) final-state interaction. We note that, if implemented as a constant amplitude, the ${}^2S_{1/2}$ FSI becomes the dominant FSI at higher energies, and is somewhat larger than its nominal NLO size for energies above 7--8 MeV, see the red curve in Fig.~\ref{fig:nlo_fsi_cmp2}. 

However, this version of the $nc$ amplitude also overshoots the sum rule, see the right panel of \cref{fig:nlo_fsi_cmp2}, presumably because it is not a unitary amplitude. The unitarity term in this amplitude is not strictly part of the NLO amplitude, but if we include it the agreement with the sum rule gets better, see the purple line in the right panel.
There is still a small violation of the sum rule. 
This is due to the fact that the zero-range implementation with the positive scattering length
supports a bound state.
Therefore the corresponding M{\o}ller operator is isometric but not unitary.
For a more detailed discussion see \cref{ssec:resultswithFSI}.
With this \({}^2S_{1/2}\) FSI with unitarity term the difference between the result with this FSI and the the curve without FSIs remains very modest, even up to energies above 10 MeV.

\section{Results based on the multiple-scattering series}
\label{ap:mss}

\begin{figure}[H]
    \centering
    \includegraphics[width=0.45\textwidth]{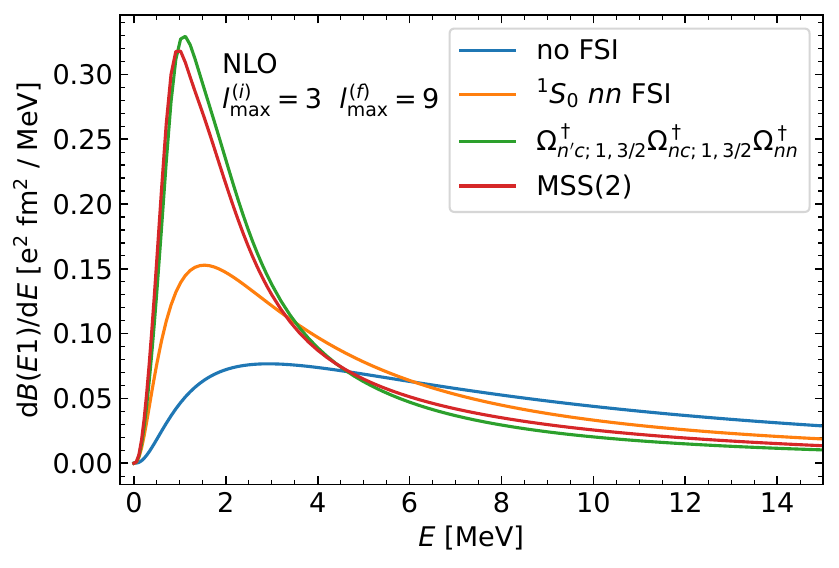}
    \caption{The NLO FREFT \(E1\) distribution is shown for different FSI treatments.
    In addition to results without FSI and only \(nn\) FSI, also a third-order MOPA result and a result based on the truncation of the multiple-scattering series at second order in the \(G_0 t_{ij}\) (MSS(2)) are shown.}
    \label{fig:mss}
\end{figure}

In addition to using solely single FSIs and the approximation strategy based on products of M{\o}ller operators (MOPA), we tried also direct truncations of the multiple-scattering series at different orders in the \(G_0 t_{ij}\).
We tested first and second order. While at second order the violation of the sum rule is small, at first order it is large.
The reason is that this method doesn't need to be unitary.
\Cref{fig:mss} shows this result in comparison with other results already presented in Sec.~\ref{ssec:resultswithFSI}.

We observe that the MSS(2) result and the result based on the product of three M{\o}ller operators are quite similar.
This might be related to the fact that if the product of three operators is multiplied out in terms of \(\Omega_{ij}^\dagger = \id + \bar{\Omega}_{ij}^\dagger\), it shares a lot of terms with the MSS(2) expression.
The terms that are present in the one approach and not in the other apparently cancel approximately.

\section{Convergence in the included partial waves}
\label{ap:pw_trunc}

The ground-state wave function has infinitely many partial-wave components indicated by the multiindices \(\Omega_i\).
Therefore, some truncation determining which partial-wave states get included is necessary.
If no FSI is applied or the FSIs do not require recouplings in their evaluation, then for each initial partial-wave state only finitely many final partial-wave states after \(E1\) breakup (and FSI) are possible. Only \(\lambda_f\), \(I_f\), and \(J_f\) can vary, but they can deviate by at most 1 from the corresponding initial quantum number, if no other restrictions are present. Therefore, there are finitely many of them.
If the evaluation of FSIs requires recoupling, the situation is a bit more complex.
For the state after the \(E1\) breakup itself the previous argumentation applies, i.e., there are only finitely many states.
Then after a recoupling, the FSI has to be evaluated with respect to the corresponding spectator.
Because the FSI is only present in a certain partial-wave channel for the two-body subsystem, \(l\), \(s\), and \(j\) of that intermediate
state are already determined. Moreover, \(\sigma\) is fixed by the recoupling, while \(J\) and \(M\) are conserved by it.
So, for each recoupling in the evaluation of a sequence of FSIs, there are only finitely many intermediate partial-wave states.
Then there is the final recoupling to the final state with respect to the core as spectator.
Since after that step there is no further FSI, which would project on certain quantum numbers, only \(\sigma\), \(J\) and \(M\) of the final state are fully determined.
In principle, for each of the finitely many considered initial states infinitively many final states are possible in this situation and another truncation is necessary.

As a consequence of these observations for the two cases, (a) no FSI / FSI with no recoupling, (b) FSI with recoupling, we decided to truncate
in the considered initial states and the considered final states. Independently of case (a) or (b) we use the same mechanism. For the intermediate states in case (b) discussed before, we apply no truncation.
Our truncation scheme is to consider only those initial states \(\Omega_i\), where \(l_i \leq \lmi \, \land \, \lambda_i \leq \lmi\) holds. Applying the condition only on \(l_i\) would already result in finitely many states, however, we prefer this more symmetric truncation.
For the final states we impose the analogous condition: \(l_f \leq \lmf \, \land \, \lambda_f \leq \lmf\).
For an equation making use of the multiindices for the partial waves we refer to \cref{eq:e1_distrib_final}.

\begin{figure}[H]
    \centering
    \includegraphics[width=0.98\textwidth]{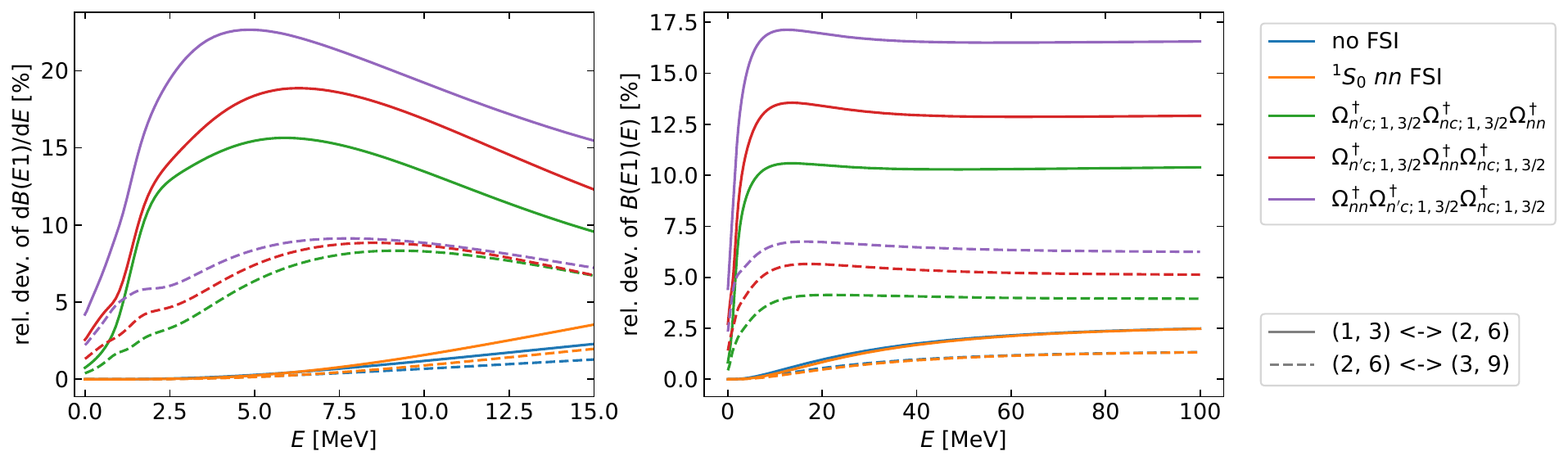}
    \caption{\textit{Left panel:} Relative deviation in percent between \(E1\) distributions obtained with different truncations in the inclusion of partial waves for the case of different FSIs included.
    The solid lines show relative deviations between distributions using \(\lmi=1\), \(\lmf=3\) and distributions using \(\lmi=2\) and \(\lmf=6\).
    The dashed lines show the relative deviations between distributions using \(\lmi=2\), \(\lmf=6\) and distributions using \(\lmi=3\) and \(\lmf=9\).
    All calculations were done at NLO.
    \textit{Right panel:} The relative deviations in percent between cumulative distributions obtained with different truncations are shown. The parameters are the same as in the left panel.
    }
    \label{fig:e1_pw_cvg}
\end{figure}

Relative deviations based on comparing the distributions with different \(\lmi\) and \(\lmf\) are shown in \Cref{fig:e1_pw_cvg} for distributions differing in FSIs.

We see in the left panel that all curves show a convergence pattern.
The solid lines, which correspond to the deviations between the lower truncation parameters, display in general larger absolute values than the dashed curves.
This is true for the curves based on a product of three M{\o}ller operators as well as for the curves based on no FSI or \(nn\) FSI only.
However, the absolute values are different.
While for the deviation between \(\lmi=2\), \(\lmf=6\) distributions and \(\lmi=3\) and \(\lmf=9\) distributions the deviations for the results using three M{\o}ller operators is below 10\%, for the \(nn\) FSI or no FSI results it is below 2.5\% in the region of \(E\) up to 15~MeV.
Based on this we estimate that the uncertainty from the partial-wave truncation for our result based on \(\Omega_{n'c}^\dagger \Omega_{nc}^\dagger \Omega_{nn}^\dagger\) is below 10\% at \(E=15\)~MeV and below 5\% for \(E < 5\)~MeV.

\bibliographystyle{apsrev4-2}
\bibliography{tex/references.bib}

\end{document}